\documentclass[aps,prd,10pt,twocolumn,superscriptaddress,floatfix,nofootinbib,amsmath,amssymb,altaffilletter,preprintnumbers]{revtex4-2}
\usepackage[colorlinks=true,pdfstartview=FitV]{hyperref}

\usepackage{graphicx}
\usepackage{multirow}
\usepackage{soul}
\usepackage{natbib}
\usepackage{amsmath}
\usepackage{fontawesome}
\usepackage{mathrsfs}
\usepackage{amssymb}
\usepackage{bm}
\usepackage{yfonts}
\usepackage{makecell}
\usepackage{color}
\usepackage{xspace}
\usepackage{url}
\usepackage{verbatim}
\usepackage{mathtools}
\usepackage{upgreek}
\usepackage{units}
\usepackage{amstext}
\usepackage[sc]{mathpazo}
\usepackage{helvet}
\usepackage{booktabs}
\usepackage{gensymb}
\usepackage{tabulary}
\usepackage{tabularx}
\usepackage{etoolbox}
\usepackage[version=4]{mhchem}
\usepackage[utf8]{inputenc}
\usepackage[dvipsnames,table]{xcolor}
\newcolumntype{Y}{>{\centering\arraybackslash}X}
\hypersetup{urlcolor=MidnightBlue,
	    citecolor=Periwinkle,
	    linkcolor=CadetBlue}
\usepackage[official]{eurosym}
\definecolor{lightgray}{rgb}{0.9,0.9,0.9}	    
\definecolor{green}{rgb}{0,0.5,0}
\definecolor{red}{rgb}{1,0,0}
\definecolor{blue}{rgb}{0,0,0.5}

\long\def\symbolfootnote[#1]#2{\begingroup%
\def\thefootnote{\fnsymbol{footnote}}\footnotetext[#1]{#2}\footnotemark[#1]\endgroup}

\newcommand{\dbd}[2]{\ifmmode \frac{\textrm{d}#1}{\textrm{d}#2}\else $\textrm{d}#1/\textrm{d}#2$\fi}
\newcommand{\pbp}[2]{\ifmmode \frac{\partial#1}{\partial#2}\else $\partial#1/\partial#2$\fi}

\DeclareMathAlphabet{\mathpzc}{OT1}{pzc}{m}{it}
 \newcommand{\eV}{\text{e\kern-0.15ex V}\xspace}

 \newcommand{\TeV}{\text{T\kern-0.1ex \eV}\xspace}

\DeclareMathAlphabet{\mathpzc}{OT1}{pzc}{m}{it}

\newcommand{\be}{\begin{equation}}
\newcommand{\ee}{\end{equation}}
\newcommand{\bea}{\begin{eqnarray}}
\newcommand{\eea}{\end{eqnarray}}

\renewcommand\({\left(}
\renewcommand\){\right)}

\def\l#1{\label{eq:#1}}

\begin{document}  

\title{Searching For Dark Matter with Plasma Haloscopes}
\preprint{FERMILAB-PUB-22-739-T}
  
\author{Alexander J. Millar}\email{amillar@fnal.gov}
\affiliation{The Oskar Klein Centre, Department of Physics, Stockholm University, AlbaNova, SE-10691 Stockholm, Sweden}
\affiliation{Nordita, KTH Royal Institute of Technology and
Stockholm
  University, Roslagstullsbacken 23, 10691 Stockholm, Sweden}
  \affiliation{Fermi National Accelerator Laboratory, Batavia, Illinois 60510, USA}

\author{Steven M. Anlage}
\affiliation{Quantum Materials Center, Physics Department, University of Maryland, College Park, MD 20742-4111 USA}

\author{Rustam Balafendiev}
\affiliation{Science Institute, University of Iceland, 107 Reykjavik, Iceland}
\author{Pavel Belov}
\affiliation{Narxoz University, Zhandossov street 55, 050035 Almaty, Kazakhstan}
\author{Karl van~Bibber}
\affiliation{Department of Nuclear Engineering, University of California Berkeley, Berkeley, CA 94720 USA}

\author{Jan Conrad}
\affiliation{The Oskar Klein Centre, Department of Physics, Stockholm University, AlbaNova, SE-10691 Stockholm, Sweden}

\author{Marcel Demarteau}
\affiliation{Physics Division, Oak Ridge National Laboratory, TN 37831}

\author{Alexander Droster}
\affiliation{Department of Nuclear Engineering, University of California Berkeley, Berkeley, CA 94720 USA}

\author{Katherine Dunne}
\affiliation{The Oskar Klein Centre, Department of Physics, Stockholm University, AlbaNova, SE-10691 Stockholm, Sweden}

\author{Andrea Gallo~Rosso}
\affiliation{The Oskar Klein Centre, Department of Physics, Stockholm University, AlbaNova, SE-10691 Stockholm, Sweden}

\author{Jon E. Gudmundsson}
\affiliation{The Oskar Klein Centre, Department of Physics, Stockholm University, AlbaNova, SE-10691 Stockholm, Sweden}
\affiliation{Science Institute, University of Iceland, 107 Reykjavik, Iceland}

\author{Heather Jackson}
\affiliation{Department of Nuclear Engineering, University of California Berkeley, Berkeley, CA 94720 USA}

\author{Gagandeep Kaur}
\affiliation{Centre for Lasers and Photonics, Indian Institute of Technology Kanpur; Kanpur, 208016, India}
\affiliation{The Oskar Klein Centre, Department of Physics, Stockholm University, AlbaNova, SE-10691 Stockholm, Sweden}

\author{Tove Klaesson}
\affiliation{The Oskar Klein Centre, Department of Physics, Stockholm University, AlbaNova, SE-10691 Stockholm, Sweden}

\author{Nolan Kowitt}
\affiliation{Department of Nuclear Engineering, University of California Berkeley, Berkeley, CA 94720 USA}

\author{Matthew Lawson }
\affiliation{The Oskar Klein Centre, Department of Physics, Stockholm University, AlbaNova, SE-10691 Stockholm, Sweden}
\affiliation{Nordita, KTH Royal Institute of Technology and
Stockholm
  University, Roslagstullsbacken 23, 10691 Stockholm, Sweden}

\author{Alexander Leder}
\affiliation{Department of Nuclear Engineering, University of California Berkeley, Berkeley, CA 94720 USA}

\author{Akira Miyazaki}
\affiliation{Department of Physics and Astronomy, Uppsala University, Uppsala, 75237, Sweden}

\author{Sid Morampudi}
\affiliation{Center for Theoretical Physics, MIT, Cambridge, MA 02139 USA}

\author{Hiranya V. Peiris}
\affiliation{The Oskar Klein Centre, Department of Physics, Stockholm University, AlbaNova, SE-10691 Stockholm, Sweden}
\affiliation{University College London, Gower Street, London, WC1E 6BT, UK}

\author{Henrik S. Røising}
\affiliation{Niels Bohr Institute, University of Copenhagen, DK-2200 Copenhagen, Denmark}

\author{Gaganpreet Singh}
\affiliation{The Oskar Klein Centre, Department of Physics, Stockholm University, AlbaNova, SE-10691 Stockholm, Sweden}

\author{Dajie Sun}
\affiliation{Department of Nuclear Engineering, University of California Berkeley, Berkeley, CA 94720 USA}
\author{Jacob H. Thomas}
\affiliation{X1, Department of Physics, Illinois Institute of Technology, Chicago, IL 60616 USA}
\author{Frank Wilczek}
\affiliation{The Oskar Klein Centre, Department of Physics, Stockholm University, AlbaNova, SE-10691 Stockholm, Sweden}
\affiliation{Center for Theoretical Physics, MIT, Cambridge, MA 02139 USA}
\affiliation{T.D.~Lee Institute and Wilczek Quantum Center, 
Shanghai Jiao Tong University, Shanghai 200240, China}
\affiliation{Arizona State University, Tempe, AZ 85287, USA}

\author{Stafford Withington}
\affiliation{Department Physics, University of Oxford, Clarendon Laboratory
Parks Road, Oxford, OX1 3PU}
\author{Mackenzie Wooten}

\affiliation{Department of Nuclear Engineering, University of California Berkeley, Berkeley, CA 94720 USA}
\collaboration{%\vspace{ex}
Endorsers\vspace{-2ex}}
%\noaffiliation
\author{Jens Dilling}
\affiliation{Physics Division, Oak Ridge National Laboratory, TN 37831}
\author{Michael Febbraro}
\affiliation{Physics Division, Oak Ridge National Laboratory, TN 37831}
\author{Stefan Knirck}
  \affiliation{Fermi National Accelerator Laboratory, Batavia, Illinois 60510, USA}
  
\author{Claire Marvinney}
  \affiliation{Physics Division, Oak Ridge National Laboratory, TN 37831}
  
\begin{abstract}
  We summarise the recent progress of the Axion Longitudinal Plasma HAloscope (ALPHA) Consortium, a new experimental collaboration to build a plasma haloscope to search for axions and dark photons. The plasma haloscope is a novel method for the detection of the resonant conversion of light dark matter to photons. ALPHA will be sensitive to QCD axions over almost a decade of parameter space, potentially discovering dark matter and resolving the Strong CP problem. Unlike traditional cavity haloscopes, which are generally limited in volume by the Compton wavelength of the dark matter, plasma haloscopes use a wire metamaterial to create a tuneable artificial plasma frequency, decoupling the wavelength of light from the Compton wavelength and allowing for much stronger signals. We develop the theoretical foundations of plasma haloscopes and discuss recent experimental progress. Finally, we outline a baseline design for ALPHA and show that a full-scale experiment could discover QCD axions over almost a decade of parameter space.
\end{abstract}
\maketitle
\tableofcontents
 \section{Introduction \label{sec:introduction}}%Of the unsolved cosmological mysteries facing fundamental physics today, dark matter (DM) may be the most pressing. 

A wide variety of astronomical observations and the modern theoretical understanding of structure formation in the Universe support the hypothesis that a new form of matter, not accounted for in our standard model of fundamental physics, supplies much of the mass of the Universe.  Indeed, what is often called the ``standard model of cosmology'' and abbreviated $\Lambda$CDM incorporates both a cosmological term $\Lambda$ and Cold Dark Matter, a hypothetical substance that interacts very feebly with ordinary (baryonic) matter and with itself.  

Because of the equivalence principle, the astronomical and cosmological evidence gives us only limited information about the nature of dark matter. The simplest hypothesis, conceptually, is that it is a very long-lived elementary particle that does not carry electromagnetic or color charge.  This particle must be abundantly produced in the early universe but then decoupled at a time when its typical velocities are non-relativistic (i.e, it is ``cold'').  

In recent years much attention has focused on axions as a candidate to provide the dark matter in our Universe.  Axions first appeared superficially in an unrelated context, namely the issue of why a parameter, $\theta$, that appears in the standard model is empirically constrained to be exceedingly small: $| \theta | \lesssim 10^{-10}$.   A ``naturally'' large value of $\theta$, i.e. $\theta \sim 1$, would introduce T violating effects - specifically, electric dipole moments of protons and neutrons - at levels that far exceed experimental constraints.  This difficulty led Peccei and Quinn~\cite{Peccei:1977hh} to propose extending the standard model to incorporate an additional (anomalous and spontaneously broken) $U(1)$ symmetry, now known as Peccei-Quinn or PQ symmetry.  Weinberg and Wilczek~\cite{Weinberg:1977ma,Wilczek:1977pj} independently observed that this proposal necessarily leads to the existence of a new light spin 0 particle, the axion, that is highly stable and interacts very feebly with ordinary matter.  Unfortunately, present-day theory is unable to predict the precise value of the axion mass $m_a$, however the requirement to solve the Strong CP problem tightly constrains the interactions of the axion relative to that mass, leading to an almost one-parameter theory.

Consideration of the production of axions in the early Universe and their subsequent evolution ~\cite{Preskill:1982cy,Abbott:1982af,Dine:1982ah,Bergstrom:2000pn,Jaeckel:2010ni,Feng:2010gw} reveals that their properties are consistent with the observed properties of cold dark matter.

Big Bang cosmology incorporating axions follows two qualitatively distinct scenarios, depending on whether an inflationary epoch intervenes between axion decoupling and the present.   If not, then it is possible -- although challenging -- to relate $m_a$ to the density of axion matter present in the Universe today ({\it post-inflationary scenario}).  If axions dominate the observed dark matter, the most recent calculations ~\cite{Kawasaki:2014sqa,Vaquero:2018tib,Klaer:2017ond,Buschmann:2019icd,Gorghetto:2018myk,Kawasaki:2018bzv,OHare:2021zrq,Buschmann:2021sdq} favor $m_a=40-180\,\mu$eV (with $m_a=65\pm 6\,\mu$eV if the spectrum of axions radiated as during the decay of topological defects is scale invariant).

On the other hand, reheating after inflation might not restore PQ symmetry. In this case, inflation essentially enforces a constant (but random) initial value $\theta_i$ within the entire observable Universe, thus providing an unknown initial condition.  This opens the possibility of axion models with significantly smaller values of $m_a$, correlated with very small values of $\theta_i$ ({\it pre-inflationary scenario}).  But if we insist on avoiding fine-tuning or anthropic reasoning, it is natural to estimate $\theta_i={\cal O}(1)$, and then we find  $10^{-1}\,\mu{\rm eV}\lesssim m_a\lesssim 100\,\mu\rm eV$.\footnote{Other options include modifying the cosmology so that $\theta$ is not a random choice between $\{0,2\pi\}$, for example Refs.~\cite{Graham:2018jyp,Takahashi:2018tdu}.}

These considerations motivate experiments to test the hypothesis that axions, with a mass in the range of ${\cal O}(10)-{\cal O}(100)\,\mu$eV, exist and constitute the cosmological dark matter.  The traditional approach to axion (and similar hidden photon~\cite{Jaeckel:2010ni,Jaeckel:2013ija,Fabbrichesi:2020wbt,Caputo:2021eaa}) dark matter searches is based on use of a resonant cavity, designed so that a resonant mode (typically the lowest transverse magnetic mode) matches the Compton wavelength of the dark matter~\cite{Sikivie:1983ip}. Whilst this is adequate for relatively low frequencies (hundreds of MHz to a few GHz~\cite{Sikivie:1983ip,Rybka:2014xca,Woohyun:2016}) the diminishing size of the cavity at high frequencies leads to a rapid deterioration in signal power. 

This problem, and the perceived promise of the axion hypothesis, has inspired a torrent of new experimental ideas.  Some proposals relevant to the indicated mass range involve using multiple or coupled cavities~\cite{Goryachev:2017wpw,Melcon:2018dba,Melcon:2020xvj,Jeong:2020cwz,Aja:2022csb};
%though these approaches may be difficult to scale to a very large number of cavities. 
others abandon the traditional cavity altogether, and substitute either a mirror, as in the case of a dish antenna ~\cite{Horns:2012jf,Jaeckel:2013sqa,Suzuki:2015sza,Experiment:2017icw,BREAD:2021tpx} or an array of large dielectric disks \cite{TheMADMAXWorkingGroup:2016hpc,Baryakhtar:2018doz, Chiles:2021gxk,Manenti:2021whp}.  (Completely different ideas come into play for ultra-small values of $m_a$.) For recent comprehensive reviews see~\cite{Irastorza:2018dyq,Sikivie:2020zpn,Adams:2022pbo}.

These approaches, broadly speaking, attempt to overcome the barrier to resonance that arises due to the mismatch between the fundamentally massless photon and a massive, non-relativistic axion by breaking translation invariance, thus providing momentum.  The leading idea of Ref.~\cite{Lawson2019}, pursued by ALPHA and discussed in detail here, is instead to provide the photon with an effective mass corresponding to its plasma frequency.\footnote{In THz regime there have been some similar concepts to make an effective massive photon quasiparticle (polariton) using condensed matter axions~\cite{Marsh:2018dlj,Schutte-Engel:2021bqm} and optical phonons~\cite{Mitridate:2020kly,Marsh:2022fmo}.}  This has the fundamental advantage that it allows wavelength matching up to the de Broglie rather than the Compton wavelength of the dark-matter axions, and thus allows use of much larger resonant systems.  

As is well known, photons acquire an effective mass inside a plasma.  No natural plasma has all the required properties ({\it viz}., cryogenic operation, low loss, and tuneability) within the frequency range of interest, but an {\it artificial} plasma, consisting of an array of thin wires~\cite{Pendry:1998}, appears to be suitable.  Notably, because the properties of wire metamaterials depend primarily on the geometry of the system, the plasma frequency can be tuned through geometric changes~\cite{Gelmini:2020kcu,Wooten:2022vpj}. These properties make wire metamaterials excellent candidate materials to implement the concept of a plasma haloscope.

The application of plasma haloscopes is not limited to the QCD axions from the spontaneously broken PQ symmetry. More generally, a plasma haloscope can address any light particles interacting with high-frequency microwaves. Physics beyond the Standard Model, such as string theory, universally predicts new scalar and pseudoscalar bosons or extra $U(1)$ gauge bosons. The former is often referred to as {\it axion-like particles} and is free from the theoretical constraint in mass and coupling of QCD axions~\cite{Masso:1995tw, Masso:2002ip, Ringwald:2012hr, Ringwald:2012cu, Arvanitaki:2009fg, Cicoli:2012sz, Jaeckel:2010ni}. The latter are hidden photons (HP) or dark photons, though paraphotons has also been used historically~\cite{Fayet:1980rr,Okun:1982xi,Georgi:1983sy,Holdom:1985ag}. The technical requirements of searching for these particles are very similar to that of axions but without needing a static magnetic field~\cite{Arias:2012az,Caputo:2021eaa}.

In this white paper we develop an analytic formalism for plasma haloscopes, proving the formulas used in Ref.~\cite{Lawson2019} and developing an overlap integral formalism applicable when the system exhibits a single mode. Using numerical simulations, we also explore the expected quality factors of wire metamaterials at low temperatures and practical tuning geometries. We then review recent experimental work by Refs.~\cite{Balafendiev:2022wua,Wooten:2022vpj} that validates our theoretical analysis. We also introduce superconducting metamaterial as an alternative candidate for a tuneable effective plasma with higher unloaded quality factor. Finally, we outline an overall design for the ALPHA project and make projections for the expected sensitivities of exploratory and full scale experiments.

\section{Analytic Formulation}
\label{general}
Here we calculate the interactions of axions with our detector, a tuneable plasma. %We will focus here on aniostropic uniaxial plasmas, as we would expect for the kinds of WM we consider. 

First, we will set up and formally solve the Maxwell equations coupled with the Klein-Gordon equation of axion (axion-Maxwell equation), estimating the microwave power produced in such a device.  Then, we further develop the single mode approximation, obtaining a convenient overlap integral. Finally, we exploit the analysis of Ref.~\cite{Balafendiev:2022wua} to find explicit expressions for rectangular plasma modes, enabling us to estimate the unloaded quality factor for a given choice of wire material.

\subsection{Solving the Axion-Maxwell Equations}
We are interested in calculating the conversion of axions to photons in a finite plasma. Throughout this work unless otherwise specified we use natural units with the Lorentz-Heaviside for electromagnetic units. The axion-Maxwell equations in a medium are given by
\begin{subequations}
\bea
\label{coulombaxion}
{\bm\nabla}\cdot {\bf D} &=& - g_{a\gamma} {\bf B}\cdot {\bm\nabla} a		\, ,
\label{eq:Maxwell-a-matter}\\
{\bm\nabla}{\bm \times} {\bf H} - \dot {\bf D}   &=&   g_{a\gamma}\({\bf B}\dot a  -{\bf E}{\bm \times} {\bm\nabla} a \)\, ,
\label{eq:Maxwell-b-matter}\\
{\bm\nabla}\cdot{\bf B}&=& 0\,,\label{eq:Maxwell-c}\\
{\bm\nabla}\times{\bf E}+\dot{\bf B}&=&0\,,\label{eq:Maxwell-d}\\
\ddot a-{\bm\nabla}^2 a +m_a^2 a &=& g_{a\gamma} {\bf E}\cdot {\bf B}.
\label{eq:Maxwell-c-matter}
\eea
\label{eq:Maxwell-x-matter}
\end{subequations} 
Here we have defined $\bf E$ as the electric field, $\bf B$ as the magnetic flux density with $\bf D$ and $\bf H$ the displacement and magnetic field strengths, respectively. We have neglected any free charges or currents as we are interested in axion-sourced fields. Neglecting the small spatial gradient of the axion, the axion field is giving by the real part of 
\begin{equation}
    a=a_0e^{-i m_a t}\,,
\end{equation}
where $a_0$ is the axion field strength coming from the axion dark matter density 
\begin{equation}
\rho_a=\frac{m_a^2a_0^2}{2}\,.
\end{equation}
The axion couples to the photon via a dimensional coupling $g_{a\gamma}$. Such a coupling is also often written as a dimensionless quantity $C_{a\gamma}$, which makes use of the fact that for the QCD axion the mass and coupling can be calculated from the axion decay constant $f_a$, numerically found to be~\cite{diCortona:2015jxo}
\begin{subequations}
\begin{eqnarray}
m_a&=&5.70(6)(4)\,  \mu{\rm eV}\,\left(\frac{10^{12}\rm\,GeV}{f_a}\right)\,,\label{eq:ma}\\
g_{a\gamma}&=&-\frac{\alpha}{2\pi f_a}\,C_{a\gamma}\,,\\%\nonumber\\
%&=&-2.04(3)\times10^{-16}~{\rm GeV}^{-1}\,\left(\frac{m_a}{1\,\mu{\rm eV}}\right)\,C_{a\gamma}\,,\label{eq:gag}\\
C_{a\gamma}&=&\frac{{\cal E}}{{\cal N}}-1.92(4)\label{eq:cag}\,,
\end{eqnarray}
\end{subequations}
where $\alpha$ is the fine structure constant, $\cal E$ is the electromagnetic anomaly and $\cal N$ is the colour anomaly (also known as the domain wall number) of the axion. For the simplest standard benchmark KSVZ and DFSZ models $|C_{a\gamma}|=1.92,~{\rm and}~0.746$ respectively~\cite{Kim:1979if,Shifman:1979if,Dine:1981rt,Zhitnitsky:1980tq}.

We will linearise the system around an external magnetic field ${\bf B_{\rm e}}$ that solves Maxwell's equations independently (its sources are left implicit). Thus, to first order in $g_{a\gamma}$ we have
\begin{subequations}
\bea
{\bm\nabla}\cdot {\bf D} &=& - g_{a\gamma} {\bf B_{\rm e}}\cdot {\bm\nabla} a		\, ,
\label{eq:Maxwell-a-matter2}\\
{\bm\nabla}{\bm \times} {\bf H} - \dot {\bf D}   &=& g_{a\gamma}{\bf B_{\rm e}}\dot a \, ,
\label{eq:Maxwell-b-matter2}\\
{\bm\nabla}\cdot{\bf B}&=& 0\,,\label{eq:Maxwell-c2}\\
{\bm\nabla}\times{\bf E}+\dot{\bf B}&=&0\,,\label{eq:Maxwell-d2}\\
\ddot a-{\bm\nabla}^2 a +m_a^2 a &=& g_{a\gamma} {\bf E}\cdot {\bf B_{\rm e}}\,,
\label{eq:Maxwell-c-matter2}
\eea
\label{eq:Maxwell-x-matter2}
\end{subequations} 
where now $\bf B$ only contains the axion-induced magnetic fields.
We will now turn our attention to a plasma that is infinite in one direction (which we will take to be the z direction), but bounded in the transverse directions.  This defines a commonly encountered ``plasma waveguide" configuration, whose analysis we will borrow~\cite{bellan_2006}. 

As we anticipate a wire metamaterial (WM), we will assume that the medium only has a plasma response in one direction, aligned with the cylinder, acting as vacuum in the other two directions. This is summarised in a an electric permittivity $\epsilon_z$.  For simplicity we will consider $\mu=1$. The symmetry of the system allows us to break up the fields into the transverse and z directions, writing
\begin{equation}
		{\bf B}={\bf B}_t+B_z \hat {\bf z}\,;\quad  {\bf E}={\bf E}_t+E_z \hat {\bf z}\,;\quad  {\bf B}_{\rm e}=B_{\rm e}\hat {\bf z}\, ,
\end{equation}
where the subscript $t$ stands for transverse. We can analyze the fields into harmonic components, assuming that the fields oscillate with angular frequency $\omega$, for which we derive  \eqref{eq:Maxwell-a-matter2} and \eqref{eq:Maxwell-d2}  
\begin{subequations}
\bea
	\({\bm\nabla}_t+\frac{\partial}{\partial z}\hat{\bf z}\){\bm \times} \({\bf B}_t+B_z \hat {\bf z}\)    &=& -i\omega \({\bf E}_t+\epsilon_z E_z\hat {\bf z} \)\nonumber\\
	&-&i\omega g_{a\gamma} a B_{\rm e}\hat {\bf z} \, ,\\
	\({\bm\nabla}_t+\frac{\partial}{\partial z}\hat{\bf z}\){\bm \times}\({\bf E}_t+E_z\hat {\bf z} \)&=&i\omega \({\bf B}_t+B_z \hat {\bf z}\)\,.
\eea
\end{subequations}
Note that the transverse curl of the transverse vectors is necessarily in the z direction, so we can divide these equations into 
\begin{subequations}
\bea
\hat {\bf z}\cdot {\bf \nabla}_t\times {\bf B}_t&=&-i\omega\epsilon_z E_z-i\omega g_{a\gamma} a B_{\rm e}\, , \\
\hat {\bf z}\cdot {\bf \nabla}_t\times {\bf E}_t&=&i\omega B_z \, ,\\
\hat {\bf z}\times  \frac{\partial {\bf B}_t}{\partial z}+{\bf \nabla}_tB_z\times \hat {\bf z}&=&-i\omega {\bf E}_t\, , \\
\hat {\bf z}\times  \frac{\partial {\bf E}_t}{\partial z}+{\bf \nabla}_tE_z\times \hat {\bf z}&=&i\omega {\bf B}_t\,.
\eea
\end{subequations}
Taking the fields' z dependence to be of the form $e^{ik_zz}$, we arrive at a closed form for ${\bf B}_t$ and ${\bf E}_t$;
\begin{subequations}
\bea
{\bf E}_t&=&\frac{1}{\omega^2-k_z^2}\({\bf \nabla}\frac{\partial  E_z}{\partial z}+i\omega{\bf \nabla}_t{\bf B}_z\times \hat {\bf z}\)\, ,\\
{\bf B}_t&=&\frac{1}{\omega^2-k_z^2}\({\bf \nabla}\frac{\partial  B_z}{\partial z}-i\omega{\bf \nabla}_t{\bf E}_z\times \hat {\bf z}\)\, .
\eea
\end{subequations}
Thus the transverse fields depend only on $E_z$ and $B_z$.  This is a consequence of the fact that the axion driving term arises only in the direction of the external B-field. Since the axion does not couple to $B_z$ mode, for modes that do bring it in geometrically we can neglect it, and focus on $E_z$, obeying
%\begin{subequations}
\bea
\frac{\omega^2}{\omega^2-k_z^2}{\bf \nabla}^2_tE_z+\omega^2\epsilon_zE_z+m_a^2g_{a\gamma}B_{\rm e}a &=&0\,\label{eq:helm} .%\\
%\frac{\omega^2}{\omega^2-k_z^2}{\bf \nabla}^2_tB_z+\omega^2B_z&=&0\, .
\eea
%\end{subequations}
For a cylindrical structure it is easiest to bring in cylindrical coordinates, yielding
\begin{equation}
	\frac{\omega^2}{\omega^2-k^2}\(r^2\frac{\partial^2 E_z}{\partial^2 r}+r\frac{\partial E_z}{\partial r}\)+r^2\omega^2\epsilon_zE_z+r^2\omega^2g_{a\gamma}B_{\rm e}a=0\,,
\end{equation} 
where we have dropped the subscript on $k \equiv k_z$.  This is solved by
\bea
	E_z&=&-\frac{ag_{a\gamma}B_{\rm e}}{\epsilon_z}+C_1J_0(ir\sqrt{\epsilon_z}\sqrt{k^2-\omega^2})\nonumber\\
	&+&C_2Y_0(-ir\sqrt{\epsilon_z}\sqrt{k^2-\omega^2})\,,
\eea
where $J_0,Y_0$ are standard Bessel functions and $C_1, C_2$ are constants.

To incorporate a bounded medium we need to solve the differential equations in both the external and plasma region and to match the solutions by applying appropriate boundary conditions. For concreteness we will consider the outside to be a conductor; other boundary conditions can be handled similarly for example including an air gap between the conductor and plasma or absorbing material. Since $Y_0(0)=-\infty$, so for the plasma region (centered, of course, on the z-axis) we use $C_2=0$. Thus the solution inside the plasma is   
\begin{equation}
	E_z=-\frac{ag_{a\gamma}B_{\rm e}}{\epsilon_z}+C_1J_0(ir\sqrt{\epsilon_z}\sqrt{k^2-\omega^2})\,.
\end{equation}
Here the first term is the solution for a homogeneous medium. It will be approached in the limit that the plasma radius is infinite.

To proceed further we assume that we are only looking for modes which will maximially couple to the axion, and so will a uniform $z$ structure (i.e., $k=0$). The transverse fields are given by
\begin{subequations}
\bea
{\bf E}_t&=&0\, ,\\
{\bf B}_t&=&\frac{i}{\omega}\frac{\partial E_z}{\partial r}\boldsymbol{ \hat{\theta}}\, .
\eea
\end{subequations}
%\subsubsection*{Conductive walls}
To find the solution outside the cylinder, we must know the relevant boundary conditions. We will assume that only the plasma itself is magnetised (i.e., $B_e=0$ outside the plasma) and  
that the plasma is surrounded by a conductor, so that the tangential component of the $E$-field (i.e, $E_z$), vanishes. Specifying the plasma-conductor boundary at a radius $R$, we see that
\begin{subequations}
\bea
	E_z&=&-\frac{ag_{a\gamma}B_{\rm e}}{\epsilon_z}+\frac{ag_{a\gamma}B_{\rm e}}{\epsilon_z}\frac{J_0(\sqrt{\epsilon_z}r\omega)}{J_0(\sqrt{\epsilon_z}R\omega)}\,,\\
	{\bf B}_t&=&-\frac{ag_{a\gamma}B_{\rm e}}{\sqrt {\epsilon_z}}\frac{J_1(\sqrt{\epsilon_z}r\omega)}{J_0(\sqrt{\epsilon_z}R\omega)}\boldsymbol{ \hat{\theta}}\, .
\eea
\end{subequations}

We are primarily interested in resonant (longitudinal) modes, where ${\rm Re}(\epsilon_z)=0$. As $J_0(0)=1$, behaviour in the centre of the medium is determined by $J_0(\sqrt{\epsilon_z}R\omega)$. As $R\to \infty$, $J_0(\sqrt{\epsilon_z}R\omega)\to\infty$, so the second term in the E-field vanishes. Similarly, $J_1(0)=0$ implies that the B-field only exists in the outer portions of the medium. Thus a sufficiency large medium has a bulk that behaves exactly the same as in the infinite medium case. However, note that this behaviour depends on $ \epsilon_z''$; smaller $ \epsilon_z''$ requires a larger $R$ for the medium to appear homogeneous. 
% In order to get an appreciation for the size of these effects, we can define
% \begin{equation}
% E_0=g_{a\gamma}{B}_{\rm e}a_0=1.3\times10^{-12}~{\rm V}/{\rm m}~\(\frac{B_{\rm e}}{10~{\rm T}}\)~
% C_{a\gamma}f_{\rm DM}^{1/2} \, .\label{eq:Ea0}
% \end{equation} 

\subsection{Signal Readout}
Of course the goal is not just to create an $E$-field, but to measure it by coupling it to an amplifier using an antenna. To discuss the signal that could be read out by an amplifier, we must have some description of the readout and losses. To look at the material losses, in the Drude model, which is typically used to describe metals,  $\epsilon$ can be written as
 \begin{equation}
    \epsilon = 1 - \dfrac{\omega_p^2}{\omega^2 + i \omega \Gamma_p}\simeq 1-\frac{\omega_p^2}{\omega^2}+i\frac{\Gamma_p \omega_p^2}{\omega^3}\,,
\end{equation}
where $\omega_p$ is the (angular) plasma frequency and $\Gamma_p$ is the inverse lifetime of the plasmon. To include antenna losses, i.e., the power extracted by an antenna which comprises the signal, we can add in an additional damping term, $\Gamma_a$ (which may be frequency dependent), giving a total dielectric constant
\begin{equation}
    \epsilon = 1 - \dfrac{\omega_p^2}{\omega^2 + i \omega \Gamma_p}+i\frac{\Gamma_a}{\omega}\,.
\end{equation}
(Such an expression implicitly assumes that the antenna couples efficiently to all parts of the system. As we will discuss below, for very large systems this can require multiple antennas.) This allows for a total loss rate $\Gamma_t$ near the plasma frequency in the medium, where 
\begin{equation}
    \Gamma_t=\Gamma_p\frac{\omega_p^2}{\omega^2}+\Gamma_a\,,
\end{equation}
We define the quality factor as 
\begin{equation}
	Q=\omega \frac{U}{P}\,.
\end{equation}
$U$ is the stored energy and $P=\dot U$ the power. The energy stored an isotropic medium with temporal dispersion is given by
\begin{equation}
    U=\frac{1}{4}\int\left ( \frac{\partial (\epsilon\omega)}{\omega}|\vec{E}|^2+|\vec{B}|^2\right)dV\simeq\frac{1}{2}\int|\vec{E}|^2 dV \,,\label{eq:U}
\end{equation}
where the latter holds exactly in the limit where the axion velocity and higher order powers of $g_{a\gamma}$ are neglected. As our medium is simply aligned (i.e., the $E$-field excited by the axion is only in the direction aligned with the wires), the isotropic formula gives the same answer as an anisotropic one. We also neglect spatial dispersion, as for uniaxial wires this exists only for momentum in the $z$-direction~\cite{Belov2002}.
%Thus we should be able to write radiated from a plasma as
% \begin{equation}
% 	P=\Gamma U\,,
% \end{equation}
%  The total loss rate should be given by the sum of the resistive losses $\Gamma_r$ and the energy absorbed by the detector $\Gamma_d$, which gives a coupling factor 
% \begin{equation}
% 	\kappa=\frac{\Gamma_d}{\Gamma_d+\Gamma_r}\,.
% \end{equation}
% The system is called ``critically coupled" when $\kappa=0.5$, or in other words the energy absorbed by the detector is equivalent to the other losses in the system.
To find the losses in the medium we can use the imaginary component of the effective dielectric constant $\epsilon_z''$ using~\cite{landau1995electrodynamics}
\begin{equation}
 %= \omega\iiint_{-d/2}^{d/2}w_{\text{wires}} dV  \\ =
P=\frac{\omega \epsilon_z''}{2}\int|E_z|^2 dV  \,.
\label{eq:pwires}
\end{equation}
Putting Eqns.~\eqref{eq:U} and \eqref{eq:pwires} together, we then find 
\begin{equation}
    Q=\frac{1}{\epsilon_z''}\,.
\end{equation}
% For a specific plasma, we can use a Drude-like model
%  \begin{equation}
%     \epsilon = 1 - \dfrac{\omega_p^2}{\omega^2 - i \omega \Gamma}\simeq 1-\frac{\omega_p^2}{\omega^2}+i\frac{\Gamma \omega_p^2}{\omega^3}~,
% \end{equation}
% where the plasmon lifetime again should be noted as including ``losses" to the detector.
Here we will treat the system as having a single loss factor, thus defining the ``loaded quality factor", in other words the quality factor of the system with an antenna system included. Conversely, the quality factor which only includes resistive losses is generally refered to as the ''unloaded quality factor". 
The signal power can then be written as 
\begin{equation}
    	P_s=\kappa {\cal G} V\frac{Q}{m_a}\rho_ag_{a\gamma}^2B_{\rm e}^2\,,\label{eq:power}
\end{equation}
where
\begin{equation}
	{\cal G}=\frac{1}{a_0^2g_{a\gamma}^2B_{\rm e}^2VQ^2}\int |{\bf E}|^2 dV\, .
\end{equation}
While we have defined ${\cal G}$ to serve a similar role to the geometry factor in a cavity haloscope, the definition and derivation of ${\cal G}$ differs conceptually.  It does not contain the overlap of the axion and photon wavefunctions directly, but is rather a normalisation of the stored energy in the resonator. We have defined $\kappa=\Gamma_a/\Gamma_t$ to take into account that only the power into the antenna is read out.

Note that near the plasma frequency $\epsilon\simeq i\Gamma_t/\omega_p$, so $Q=1/|\epsilon|$, and after some rearranging we reproduce the formula for the ${\cal G}$ in Ref.~\cite{Lawson2019} for $m_a=\omega_p$, 
\begin{equation}
	{\cal G}=\frac{|\epsilon_z|^2}{a_0^2g_{a\gamma}^2B_{\rm e}^2V}\int |{\bf E}|^2 dV\, .% \\
	%&=3\times 10^{-22}\, {\rm W}\frac{V}{0.3\, {\rm m}^3}\frac{m_a}{120 \, \mu{\rm eV}}\frac{Q}{10^3}\frac{\rho}{0.45\,{\rm GeV}/{\rm cm}^3}\(\frac{B_{\rm e}}{10\,{\rm T}}\)^2C_{a\gamma}^2\,, 
\end{equation}

So far, our derivation has been for axion dark matter. However, plasma haloscopes are also very sensitive to hidden photons (HP)~\cite{Gelmini:2020kcu}. HPs are a massive U(1) gauge boson which mixes with the visible photon. This interaction is described to lowest order by the Lagrangian within the propagation basis (i.e. the mass basis) as~\cite{Arias:2012az, Fabbrichesi_2021, An_2015_xe10,Holdom:1985ag}
\begin{equation}
\label{eq:lagrangian}
    \begin{aligned}
    \mathcal{L} = -\frac{1}{4} \tilde F_{\mu\nu} \tilde F^{\mu\nu} 
        - \frac{1}{4}\tilde F'_{\mu\nu}\tilde F'^{\mu\nu} &
        + \frac{\chi}{2}\tilde F_{\mu\nu}\tilde F'^{\mu\nu} \\
        + \frac{{m_X}^2}{2}\tilde X_{\mu}\tilde X^{\mu}
        + e J^\mu\tilde A_\mu \, ,
        \end{aligned}
\end{equation}
where the dark vector field is $\tilde X_{\mu}$, the dark photon tensor $\tilde F'_{\mu\nu}$, $\chi$ is the mixing coupling constant, $m_X$ is the mass of the dark photon, $e$ is the electric charge, $J_{\mu}$ is the current density vector, and $\tilde A_{\mu}$ is the electromagnetic four-potential. As shown in Refs.~\cite{Gelmini:2020kcu,Caputo:2021eaa} the resulting power generated in a plasma halsocope can be found by a simple substitution in~Eqn.~\eqref{eq:power}
\begin{equation}
    	P_s^{\rm HP}=\kappa {\cal G} VQ\rho_X\chi^2m_X\langle \cos^2\theta\rangle_T\,,\label{eq:angle}
\end{equation}
where $\theta$ is the angle between the wires (the $z$-direction) and the HP polarisation and
\begin{equation}
    \langle \cos^2\theta\rangle_T=\frac{1}{T}\int \cos^2\theta\,,
\end{equation}
is the time average of this angle over the measurement (or measurements, if multiple measurements are taken of the same frequency). Note that the power produced does not depend on the magnetic field: this means both that HP experiments do not need an expensive magnet, and also that vetos based on the signal behaviour when the magnetic field is turned off necessarily also veto HP signals~\cite{Gelmini:2020kcu}. As, aside from these subtleties, the same treatment can be made for axions and HP dark matter, we will focus on the former for clarity.

\subsection{Eigenmode Calculation}
\begin{figure}[tb]
\includegraphics[width=1\linewidth]{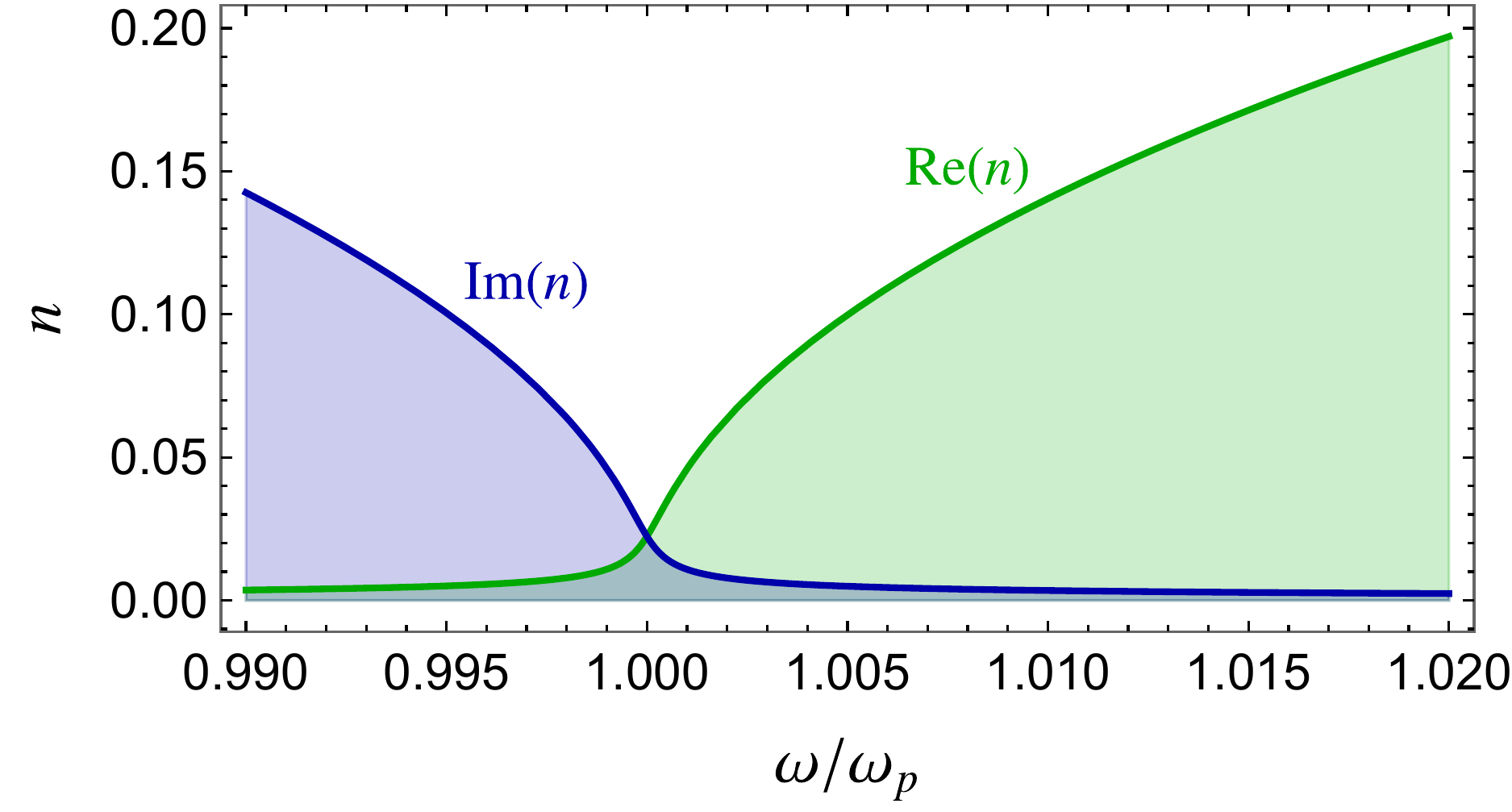}
    \caption{Real (green) and imaginary (blue) parts of the refractive index $n$ as a function of frequency $f$ around the plasma frequency $f_p$. For illustrative purposes, we have chosen a plasma with  $\Gamma_p=10^{-3}\omega_p$, i.e., a lifetime of 1000 cycles.}
    \label{fig:ref}
\end{figure}

Unlike traditional cavity haloscopes, plasma haloscopes may not have a single mode, with the shape of the $E$-field generated in the system changing with the losses and system size. As discussed above, for very large systems the device will behave almost as an infinite plasma, and there will be no coherent modes in the absence of an axion. The reason for this behaviour is the losses: unlike a cavity which is filled with air, the resistive wires lead to an imaginary component of the refractive index. For an example we can look at $n=\sqrt{\epsilon\mu}$ near the plasma frequency in Fig.~\ref{fig:ref}. At the plasma frequency the real and imaginary part of $n$ are equal, with $\epsilon_z\simeq i\Gamma_p/\omega $. Because of this traveling waves decay with a characteristic length $\sim \sqrt {\omega/\Gamma_p}\lambda_c$. This both limits the length scales over which energy can be transported, and inhibits the formation of standing waves. For systems larger than this scale, waves will decay before they travel from one wall to another.

\begin{figure}[t!]
\centering 
  \includegraphics[trim = 0mm 0mm 0mm 0mm, clip, width=0.48\textwidth]{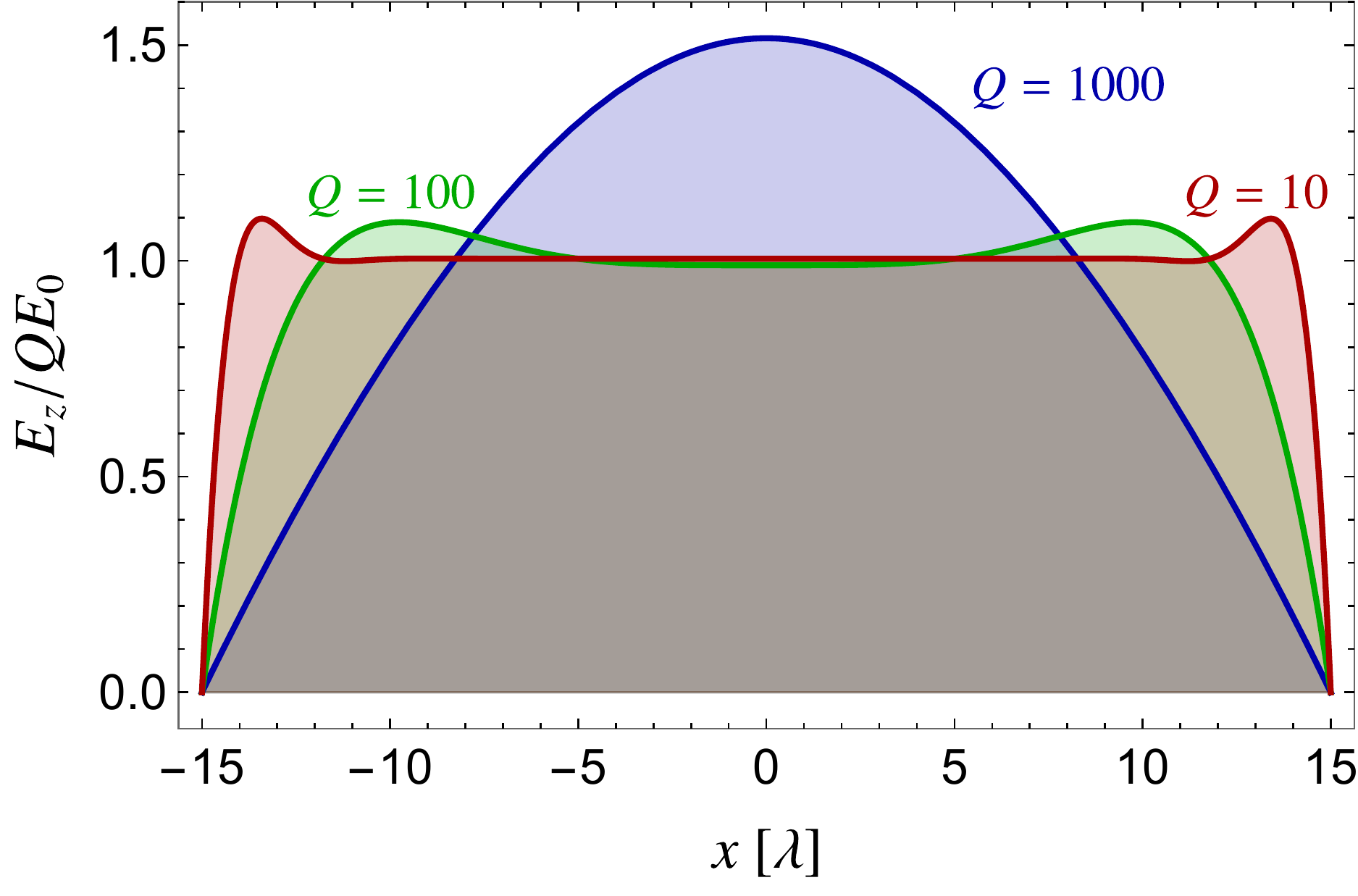}
\caption{Scaled $E$-field $E/QE_0$ for plasma haloscopes operating at their resonance frequency (frequency of maximum ${\cal G}$). To show how the mode structure changes as a function of quality factor we show $Q=10,10^2,10^3$ plasmas in a 15 wavelength cavity, depicted in red, green and blue respectively. As the decay length $\sqrt Q \lambda_c$ becomes comparable to the size of the cavity a standing cavity mode forms, which would allow single antenna readout.}
\label{fig:Qs}
\end{figure}

However, for smaller systems then the boundary conditions will influence the entire volume, giving a model structure of a cavity standing wave. We show the transition in Fig.~\ref{fig:Qs} for $Q=10,10^2,10^3$ plasmas in a 15 wavelength cavity, in each case with $\omega=\omega_p$.  If the losses are predominantly radiative or in the cavity walls, they should decrease as the system is made larger, avoiding the loss dominated regime except when resistive losses in the wires take over.

We can show explicitly that in the high-Q limit cavity modes form. In these cases, it makes sense to connect our formalism to the overlap integral approach traditionally used in cavity haloscopes. The Helmholtz-like equation~\eqref{eq:helm} can be solved for a general geometry by treating the dielectric constant $\epsilon_z$ as real and expanding in modes. To be allowed to do so requires that $\epsilon_z'>\epsilon_z''$, or 
\begin{equation}
    1-\frac{\omega_p^2}{\omega^2}\gtrsim\Gamma_p\,.
\end{equation}
Neglecting possible momentum $k$ in the $z$-direction, we can write down a heuristic expression for the resonance frequency for a mode in a more general geometry with some transverse momentum $k_t$ with
\begin{equation}
    \omega=\sqrt{\omega_p^2+k_t^2}\simeq \omega_p+\frac{k_t^2}{2\omega_p}\,,
\end{equation}
so 
\begin{equation}
    k_t\gtrsim \sqrt{\omega_p\Gamma_p}\,.
\end{equation}
For the lowest order mode in a resonator with transverse dimensions given by $\sim d$, $k_t\sim 1/d$, meaning that a single moded calculation will be appropriate when
\begin{equation}
    d\lesssim \lambda_c\sqrt{Q_p}\,.
\end{equation}
where $\lambda_c$ is the Compton wavelength at the plasma frequency, and $Q_p=\omega_p/\Gamma_p$. While we have neglected geometry specific ${\cal O}(1)$ factors, this result matches the condition that we saw numerically with the full equation, and estimates based on decay lengths in the medium.

Following the analysis of Ref.~\cite{Melcon:2018dba} we can decompose the electric field in the cavity as a series of modes $\mathcal{E}_i$,
\begin{equation}
    E_z({\bf x})=\sum_iE_i\mathcal{E}_i({\bf x})\,,
\end{equation}
 which are normalised to the cavity volume $V$ so that 
 \begin{equation}
     \int dV\mathcal{E}_i\mathcal{E}_j=V\delta_{ij}\,.\label{eq:ortho}
 \end{equation}
 In the absence of an axion, the modes must satisfy a free Helmholtz equation, so for a mode with frequency $\omega_i$
 \begin{equation}
     \nabla^2_t\mathcal{E}_i=\epsilon_z'(k_z^2-\omega_i^2)\mathcal{E}_i\,.\label{eq:eigen}
 \end{equation}
We can now rewrite~\eqref{eq:helm} as
\bea
    	&&\frac{\omega^2}{\omega^2-k_z^2}{\bf \nabla}^2_t\sum_iE_i\mathcal{E}_i+\omega^2\epsilon_z'\sum_iE_i\mathcal{E}_i+i\omega\Gamma_t \sum_iE_i\mathcal{E}_i\nonumber\\
    	&&=-m_a^2g_{a\gamma}B_{\rm e}a\,.
\eea
Using Eqns.~\eqref{eq:ortho} and~\eqref{eq:eigen}, we can remove the sums and write
\begin{equation}
    \omega^2\epsilon_z'	\frac{k_z^2-\omega_i^2}{\omega^2-k_z^2}E_i+\omega^2\epsilon_z'E_i+i\omega\Gamma_t E_i  =-m_a^2g_{a\gamma}\bar B_{\rm e}\sqrt{{\cal C}}a\,,
\end{equation}
where we have defined $\bar B_{\rm e}$ as the average value of $B_{\rm e}({\bf x})$ and 
\begin{equation}
    {\cal C}=\frac{1}{\bar B_{\rm e}^2V^2}\left(\int dV B_{\rm e} {\mathcal E}_i\right)^2\,.\label{eq:formfactor}
\end{equation}
Noting that the axion can be approximated as being spatially homogeneous with $\omega=m_a$  we can then write
\begin{equation}
    E_i=\frac{-g_{a\gamma}\bar B_{\rm e}\sqrt{{\cal C}}a}{\epsilon_z'\frac{m_a^2-\omega_i^2}{m_a^2-k_z^2}+i\frac{\Gamma_t}{\omega}}\,
\end{equation}
which gives, on resonance,
\begin{equation}
    |E_i(m_a=\omega_i)|=\frac{m_a g_{a\gamma}\bar B_{\rm e}\sqrt{{\cal C}}a}{\Gamma_t}\,.
\end{equation}
From Ref.~\cite{Melcon:2018dba} one can show that $Q=\omega/\Gamma_t$, leaving the power into the antenna as
\begin{equation}
    P_s=\Gamma_aU=\kappa\frac{Q}{m_a}g_{a\gamma}^2\bar B_{\rm e}^2{\cal C}V\rho_a\,.
\end{equation}
While this is a very similar form to Eqn.~\eqref{eq:power}, it holds only when the system forms standing wave cavity modes. To connect the two, even when $\omega_i> \omega_p$, we can identify that the calculations agree when
\begin{equation}
    {\cal C}= {\cal G}\,,
\end{equation}
which should hold when the system is single moded. 

\begin{figure}[t!]
\centering 
  \includegraphics[trim = 0mm 0mm 0mm 0mm, clip, width=0.48\textwidth]{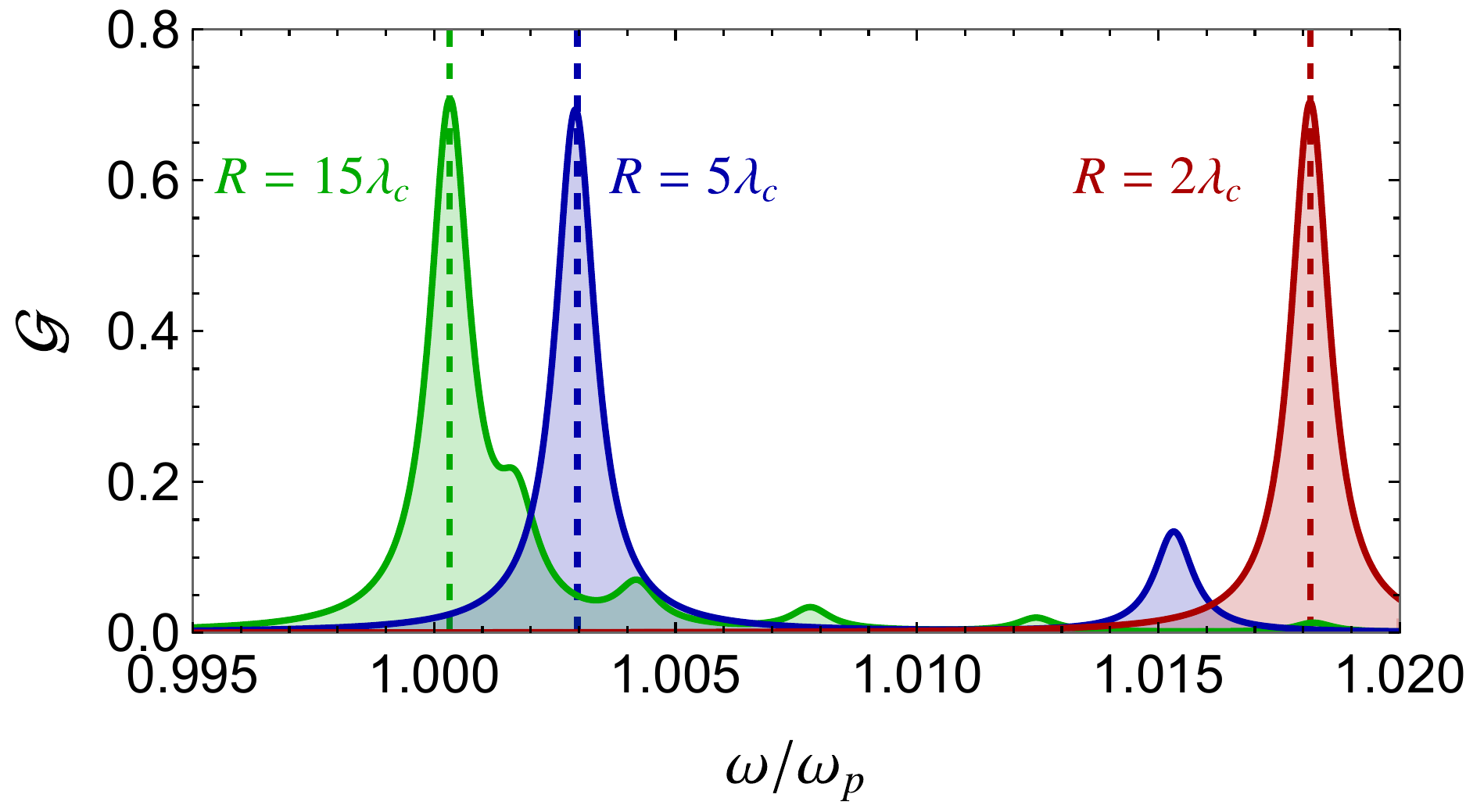}
    \includegraphics[trim = 0mm 0mm 0mm 0mm, clip, width=0.48\textwidth]{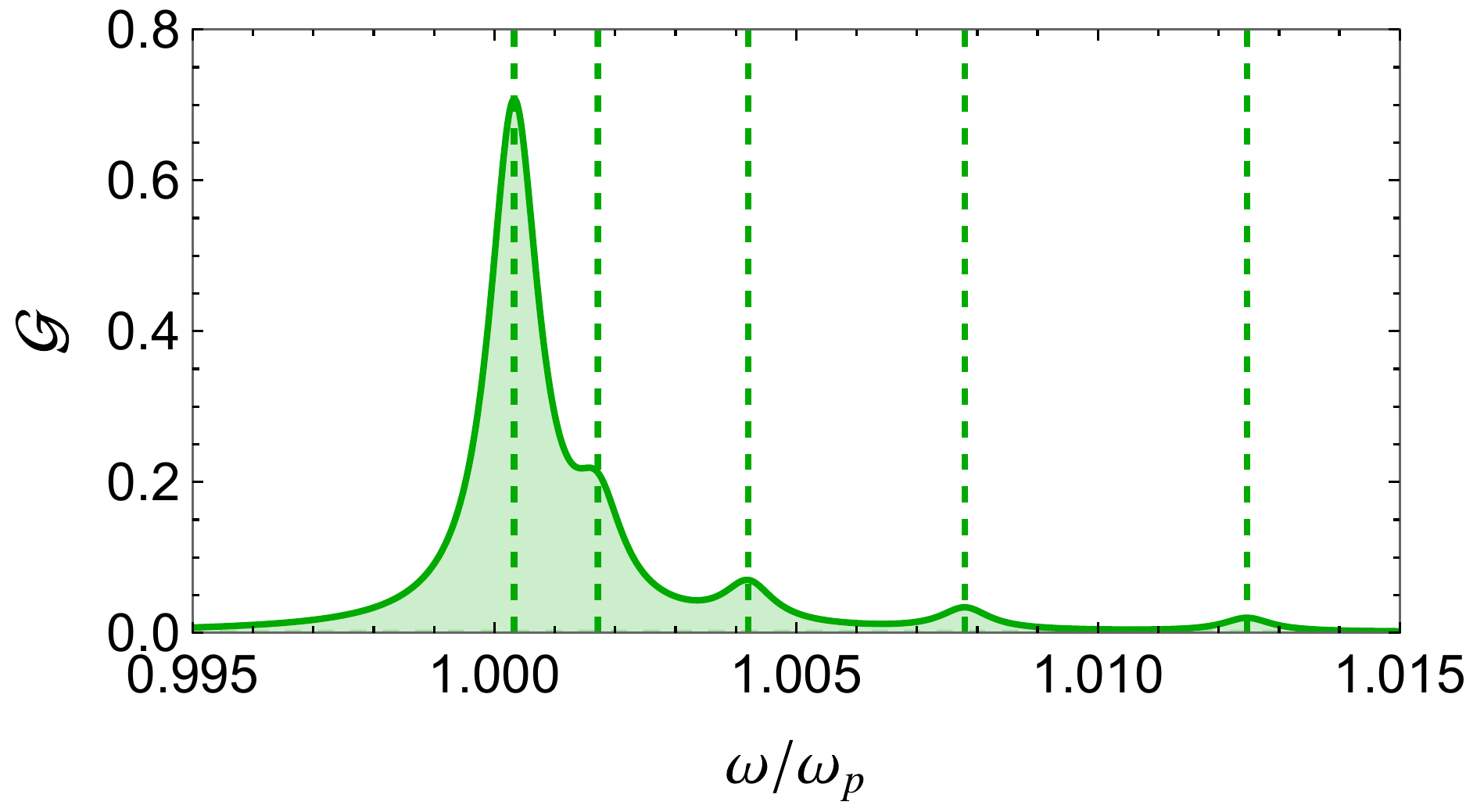}
\caption{{\bf Top}: Geometry factor $\cal G$ for 2, 5 and 15 wavelength radius cavities, depicted in red, blue and green respectively. The dashed lines indicate the lowest eigenmode of the system without an axion. The plasma is chosen to have $\Gamma_p=10^{-3}\omega_p$. {\bf Bottom}: A closer look at the eigenmodes of a 15 wavelength radius cavity. The dashed lines correspond to the lowest eigenfrequencies.}
\label{fig:nogapG}
\end{figure}

It is instructive to look at the ``high Q" regime, where $\lambda \geq R$ at the plasma frequency. In this case, readout should be possible at the boundary. To show some examples, we can take the same plasma as discussed above, with $\Gamma=10^{-3}\omega_p$. By varying the size of the cylinder, we can see the frequency for which the energy stored in the device is maximised (i.e., $\cal G$ is large). We show 2, 5, and 15 wavelength radius cavities in the top panel of Fig.~\ref{fig:nogapG}. As $|\lambda/\lambda_c|\simeq 30$ at the plasma frequency each example shows a similar profile. We see that even when the wavelength is larger than the plasma size at the plasma frequency, by moving slightly off the plasma resonance we still get an large $\cal G$ (though of course smaller devices store less power overall).

By comparing the resonant frequencies with the eigenfrequencies solved in the absence of the axion (Eqn.~\eqref{eq:eigen}) we can see if solving the equations in the absence of an axion predicts the correct resonant frequency. In the high $Q$ limit we see that the strongest signal comes when one is matching the existing eigenmodes of the system.

The calculated $\cal G$ for the fundamental mode of each setup closely matches that expected from an eigenmode analysis using Eqn.~\eqref{eq:formfactor}, with the most significant deviation coming from the $R=15\lambda_c$ case (${\cal G}=0.71$ compared to ${\cal C}=0.69$). For smaller cavities the differences between the two calculations are negligible (less than 0.1\% for $R=5\lambda_c$). For the smaller cavities, the higher order modes are also in good agreement between the two formalisms. However, as shown in the bottom panel of~Fig.~\ref{fig:nogapG} when the modes begin to overlap the simple eigenmode calculation (which assumes that the mode structure is not affected by losses) breaks down. In this case we show the cavity with $R=15\lambda_c$ in more detail, which has been chosen so that the diameter of the cavity matches the decay length of the plasma. In this case, there is overlap between the modes, which leads to significant deviations in power produced by the higher modes. For example, the next highest mode has ${\cal G}=0.21$, almost a factor of two higher than the ${\cal C}=0.13$ expected from neglecting losses. Thus when the modes become crowded or the wavelength becomes close to the single moded limit one must take care to ensure that the axion coupling to the various modes is taken into account properly. 

\subsection{Example: The Rectangular Resonator}
The geometries currently being used for prototyping, for example in Refs.~\cite{Wooten:2022vpj,Balafendiev:2022wua} are rectangular, rather than cylindrical, arrays. Here we explore the modes of a rectangular cavity with an aniostropic plasma.

\subsubsection*{TM Modes} 
 In principle, for transverse magnetic (TM) modes one can solve the differential equation:
	\begin{equation}
	\frac{\omega^2}{\omega^2-k^2}\(\frac{\partial^2 E_z}{\partial^2 x}+\frac{\partial^2 E_z}{\partial^2 y}\)+\omega^2\epsilon_zE_z+\omega^2g_{a\gamma}B_{\rm e}a=0\,.
\end{equation} 

Unfortunately, such a differential equation is difficult to solve analytically. However, as discussed above in the high $Q$ limit we can use an eigenmode expansion.  To find the eigenmodes in the high Q-regime we can neglect losses, finding that for the TM$_{l,m,n}$ mode in a rectangular array of dimensions $L_x,L_y,L_z$ \footnote{The TM$_{l,m,0}$ modes form a special case in terms of normalisation and form factors. As our primary interest is not in modes with momentum in the $z$ direction we take the normalisation and form factors from the TM$_{l,m,0}$ case to avoid dividing by zero.}
\begin{subequations}
\begin{align}
	\mathcal{E}^{l,m,n}=2\sin(k_xx)\sin(k_yy)\cos(k_z z)\,,
	\label{eq:TM110}\\
	\lambda_{l,m,n}=\left (\frac{l\pi}{L_x}\right)^2+\left (\frac{m\pi}{L_y}\right)^2+\epsilon_z\left (\frac{n\pi}{L_z}\right)^2\,,
\end{align}
\end{subequations}
where $k_x=l\pi/L_x$, $k_y=m\pi/L_y$, $k_z=n\pi/L_z$. We can see that the coupling of the modes to the axion is given by
\begin{equation}
    C_{l,m,0}=\frac{64}{\pi ^4 l^2 m^2} \sin ^4\left(\frac{l\pi 
   }{2}\right) \sin
   ^4\left(\frac{m\pi  }{2}\right)
   \,.
\end{equation}
with the $TM_{110}$ mode having the strongest coupling $C_{110}=0.66$.

This analytic formalism was found to be in good agreement with numerical simulations, as shown in Ref.~\cite{Balafendiev:2022wua}, with less than 1\% discrepancy in the resonance frequencies for larger systems.
One concern with a plasma haloscope inside a conductive cavity would be the crowding of the desired fundamental mode with higher order modes.
With these eigenvalues, we can estimate the dimensions for when mode crowding becomes a problem. In each case we must demand that the modes are separated by roughly a quality factor for the modes to be distinguishable, i.e., $\Delta \omega\gtrsim \omega/Q$. For the cavity height $L_z$, we can see that between the lowest mode ($n=0$) and next heights mode ($n=1$) the mode separation is given by
\begin{equation}
	\Delta\omega^2=2\omega\Delta\omega=\frac{\pi^2}{L_z^2}\,,
\end{equation}
which gives 
\begin{equation}
	L_z\lesssim\frac{\pi}{\sqrt 2}\frac{\sqrt Q}{\omega}\,.
\end{equation}
We can make a similar estimate for the length and width. Treating them as equal, $L_x=L_y$, we can see that neglecting losses the eigenfrequencies are given by solving
\begin{equation}
	\frac{L_x}{\pi}\simeq \frac{\sqrt{l^2+m^2}}{\sqrt{\omega^2-\omega_p^2}}\,.
\end{equation}
The lowest mode is the TM$_{110}$ mode, with the TM$_{120}$ and TM$_{210}$ modes being degenerate and next highest order. A similar criterion of $\Delta\omega\gtrsim\omega/Q$ leaves
\begin{equation}
	L_x\lesssim\sqrt{\frac{3\pi}{2}}\frac{\sqrt Q}{\omega}
\end{equation}  
Thus we see that the maximum usable volume before worrying about mode crowding of the TM modes is approximately $Q^{3/2}\lambda_c^3$, where $\lambda_c$ is the Compton wavelength. 
\subsubsection*{TE Modes} 
While the axion does not couple to transverse electric (TE) modes, they will exist if the system is inside a cavity and must be discriminated against. As, by definition, TE modes have $E_z=0$, they are unaffected by the aniostropic medium. Thus we just have regular TE mode:
\begin{subequations}
\begin{eqnarray}
	B_z^{l,m,n}=2\cos(k_xx)\cos(k_yy)\sin(k_zz)\,,\\
	\lambda_{l,m,n}=\left (\frac{l\pi}{L_x}\right)^2+\left (\frac{m\pi}{L_y}\right)^2+\left (\frac{n\pi}{L_z}\right)^2\,,
\end{eqnarray}
\end{subequations}
with resonance condition 
\begin{equation}
	\left (\frac{l\pi}{L_x}\right)^2+\left (\frac{m\pi}{L_y}\right)^2+\left (\frac{n\pi}{L_z}\right)^2=\omega^2\,.
\end{equation}
As this will behave essentially like a large cavity the TE modes near our target frequency will be very high order. To give some sense of the spacing, if the cavity consists of equal sides of length $N\pi/\omega$ then the difference between the TE$_{N,0,1}$ and TE$_{N,1,1}$ modes is given by
\begin{equation}
	\Delta \omega \simeq \frac{\omega}{N^2}\,.
\end{equation}
for large $N$. If mode mixing or avoided mode crossings occur, this could significantly interfere with larger devices and become the most stringent limit on the size of a haloscope. One can also avoid or modify the TE modes to avoid interference. There are several possibilities for such a design. If wires are placed perpendicularly to the existing uniaxial array the TE mode frequencies can be shifted up to a higher frequency, avoiding the issue. 

One could also design cavity walls that also discriminate against TE Modes. One such example would be a photonic band gap (PGB) cavity~\cite{10.1007/978-3-319-92726-8_8}, which would use tighter spaced rods to enforce reflective boundary conditions for modes aligned with the rods. One may also be able to forgo walls altogether in favour of absorbtive material. As the refractive index in the medium is less than one, vacuum itself is reflective. As long as the radiative losses of the medium to the absorptive walls is lower than the resistive losses in the wires, which should hold for sufficiently large systems, there would not be a significant degradation in $Q$.

\subsection{Resistive Losses}\label{sec:losses}
As discussed above, a key parameter for both the strength of the resonance and the mode structure is the quality factor. The limited lifetime of a resonator is captured by its quality factor $Q$. To calculate the losses, we must consider the resistivity of the materials used in the system. %We will consider the fundamental TM110 mode, though our analysis would work for any of the modes. 

 For a rectangular array of wires, with a wire radius $r$ and an inter-wire spacing in each direction perpendicular to the wires of $a$ and $b$, the plasma frequency  is given by~\cite{Belov:2003} 
\begin{equation}
\label{eq:plasmafreq}
    \frac{\omega_p^2}{c^2} = \dfrac{2\pi/s^2}{\log\left(\dfrac{s}{2\pi r} \right)+F(u)}~, 
\end{equation}
where $s=\sqrt{ab}$, $u=a/b$ and
\begin{equation}
    F(u)=-\frac12 \log u+\sum_{n=1}^\infty\left(\frac{\coth(\pi n u)-1}{n} \right)+\frac{\pi u}{6} \, ,
\end{equation}
allowing for $\nu_p \sim$ GHz when $s \sim$ cm spacing.

For clarity in this section we will depart from natural units, leaving explicit factors of $c$. The plasmon inverse lifetime $\Gamma_p$ is derived in Ref.~\cite{Maslovski2009} 
\be{}
\Gamma_p = \frac{Z_w}{L_w}%\sqrt{\epsilon_0\mu_0}
\,,\\
\l{xi}
\ee{}
with $Z_w$ and $L_w$ being the wire impedance and inductance per unit length, respectively. For a square array of wires ($a=b$) with circular cross-sections, these quantities can be calculated from the magnetic permeability $\mu$ and conductance $\sigma$ of a WM as~\cite{Maslovski2009,Olyslager2005} 

\begin{subequations}
\begin{align}
L_w &= \frac{\mu_0}{2\pi}({\log{\frac{a}{2\pi r}} + F}) \,,\\
Z_w &= \text{Re}\left(\frac{\sqrt{-i\omega\mu}}{2\pi r\sqrt{\sigma} }\frac{J_0 (r\sqrt{-i\omega\mu \sigma})}{J_1 (r\sqrt{-i\omega\mu \sigma})}\right)\,.\label{LwZw}
\end{align}
    \end{subequations}

To maximise the signal strength $Q$ should be as high as possible. To minimise resistive losses, the wire radius should be carefully chosen. One might expect that, similar to power lines used for electricity distribution, thicker wires result in lower resistive losses. However, one must be mindful to stay within the axion's finite Compton wavelength.  
%\Akira{Is the analogy to power lines valid? The thing is frequency dependence of the skin depth as described in the following. AC power 50 Hz or DC power lines are different from the 10 GHz frequency range. What does matter here is current density that is a function of skin depth and surface area.}

To mimic a large plasma haloscope, we treat the system as infinitely large (i.e., resonant frequency $\omega_{\rm res}=\omega_p$ and $Q=Q_p$). We can then solve for an optimal wire radius for a given plasma frequency. To do so, we can rewrite $Q$ explicitly as~\cite{Balafendiev:2022wua}
\begin{equation}
    Q\simeq\frac{\omega_{p}}{c\Gamma_p}=\frac{\mu_0 r\sqrt{\omega_p\sigma\mu^{-1}}}{{\rm Re}\left[\sqrt{-i}\frac{J_0 (r\sqrt{-i\omega_p\mu \sigma})}{J_1 (r\sqrt{-i\omega_p\mu \sigma})}\right]}\left(\log{\frac{a}{2\pi r}}+F\right)\,.\label{eq:Q1}
\end{equation}
We will focus on the regime where $r\gg 1/\sqrt{\omega\sigma\mu}$. This occurs when the skin depth $\delta=\sqrt{2/\omega\sigma\mu}$ is much smaller than the radius of the wires. As at 10\,GHz room temperature copper has a skin depth $\delta=0.6\,\mu$m, this condition should hold true for any reasonable wire radius. In this case currents only form in a thin shell around the outside of the wire and we can simplify Eqn.~\eqref{eq:Q1} by using
\begin{equation}
\lim_{\alpha\to\infty}\frac{J_0 (\alpha\sqrt{-i})}{J_1 (\alpha\sqrt{-i})}=i\,.
\end{equation}
All together, this gives a simple expression for the quality factor~\cite{Balafendiev:2022wua}
\bea
    Q\simeq2\frac{\mu_0}{\mu}\frac{r}{\delta}\left(\log{\frac{a}{2\pi r}}+F\right)\,.\label{eq:Q2}
\eea
Thus $Q$ is determined by two main aspects. The first being the ratio of radius $r$ with the skin depth $\delta$ (up to the relative permeability, which for copper is close to unity). The second is a geometric factor due to the inductance $L_w$, which regulates $Q$ at high frequencies. This is because $a$ is a function of $r$ and $\omega_p$, falling with an increasing $\omega_p$, being explicitly given by solving Eqn.~\eqref{eq:plasmafreq}
\begin{equation}
a=\frac{2\sqrt{\pi}}{\omega_p } {\rm W}_0\left(\frac{e^{2F}}{\pi r^2\omega_p^2}\right)^{-1}\,,    
\end{equation}
where ${\rm W}_0(x)$ is the principle branch Lambert W function.

\begin{figure}[tb]
\includegraphics[width=1\linewidth]{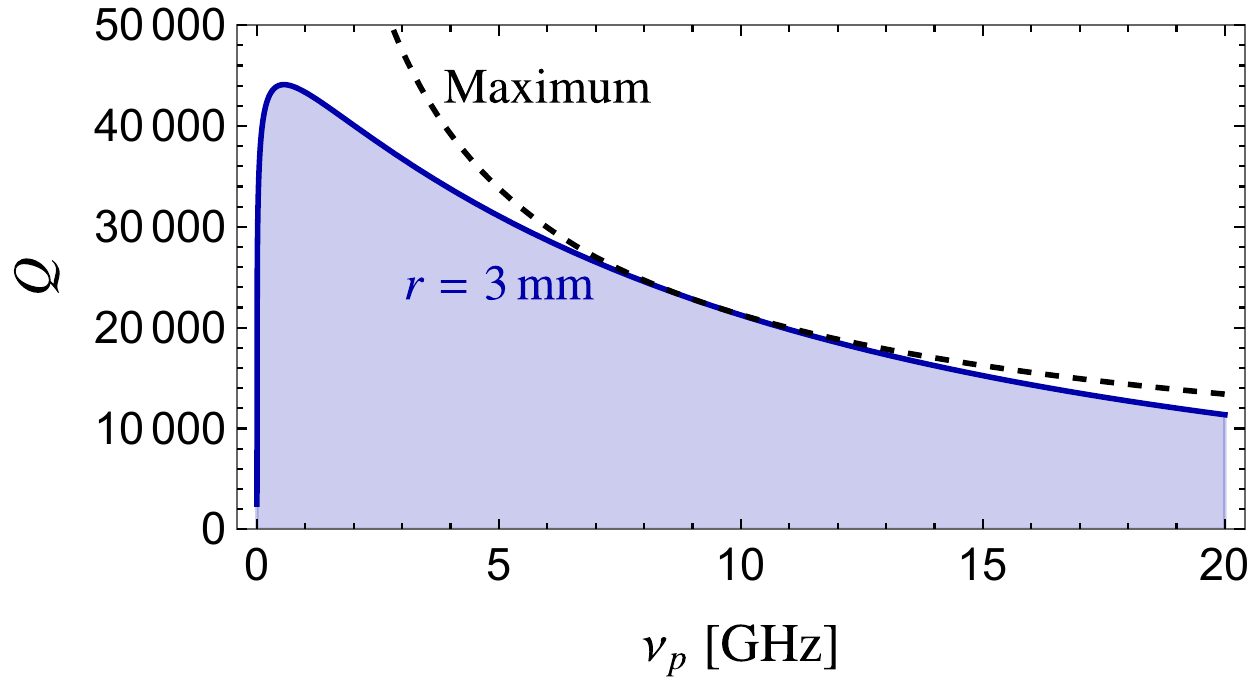}
    \caption{Unloaded quality factor $Q$ as a function of the plasma frequency $\nu_p$ for an infinite copper wire metamaterial at cryogenic temperatures (i.e., resistivity limited by the anomalous skin effect). We show wires with radius $r=3$~mm in blue with the black dashed line showing the maximum possible $Q$ for a given $\nu_p$. The period of the wires $a$ is adjusted so that the plasma frequency of the system is given by $\nu_p$.}
    \label{fig:radius}
\end{figure}

In Ref.~\cite{Balafendiev:2022wua} it was found numerically that the maximum quality factor for a given frequency occurs when $r\simeq\lambda_c/11$ assuming a constant conductance. Here, we also take into account the anomalous skin effect, where the skin depth at high frequencies in a cryogenic environment is limited to \cite{Kittel2004}
\begin{equation}
    \delta=\left[\frac{\sqrt 3 c^2 m_e v_{\rm F}}{8\pi^2\omega n e^2}\right]^{1/3}\,,
\end{equation}
where $m_e$ is the electron mass, with charge $e$ and $v_{\rm F}$ is the Fermi velocity of the metal.
Similar to Ref.~\cite{Balafendiev:2022wua} we numerically find the optimal wire radius for cryogenic copper, showing the maximum $Q$ for a given plasma frequency $\nu_p$ in the dashed line in Fig.~\ref{fig:radius}. We also show for comparison $3$\,mm wires in blue, which should be optimal around 10\,GHz. As the wire thickness approaches $\lambda_c$ the treatment of the wires as essentially 1-D objects becomes problematic, and would require changes to the effective medium description. Because of this, fine tuning the wire size should be done with full scale numerical simulations.

We can see that at cryogenic temperatures an unloaded $Q>10^4$ is possible for many frequencies under ideal conditions, even using copper. It is not until $\nu_p>30$\,GHz that the maximal $Q$ falls below this range, even then remaining high. While careful machining and design will be required to reach such high quality factors, there does not seem to be any problem in principle. One could also consider using superconducting thin wires, as we discuss in Section~\ref{sec:superconductors}, which should be capable of even higher $Q$. Such a system would be particularly interesting above 30\,GHz as the anomalous skin effect limits $Q$.

\section{Numerical Simulations \label{sec:numerical}}
Numerical simulation is a crucial step toward realistic design and engineering of the plasma haloscope. 
Simulations can test the analytical formalism and, more importantly, quantify limitations of the analytical model and go beyond it.
%Analytic formalism of the metamaterial structure may be a great way to gain insight into the overall behavior of the system, but it inevitably overlooks some finer details and interactions. 

Two important issues are that our analytical formulation becomes less reliable for very thick wires, where the uniaxial approximation breaks down; and that the average medium approach is not universally applicable.  Simulations (and physical measurements) are necessary to determine its domain of quantitative validity and semi-quantitative relevance. Here we review simulation work reported in Ref.\,\cite{Balafendiev:2022wua} using the Frequency Domain Solver in CST MWS 2020, and put it in context. We also demonstrate a new tuning scheme which achieves $30\%$ tuning without significant loss in volume. 

Throughout this section we will adopt the $10 \times 10$ wire array of $r=1$\,mm copper wires with a period of $a=10$\,mm studied in Ref.~\cite{Balafendiev:2022wua} as a benchmark and explore the mode structure, quality factor and possible tuning schemes. 
\subsection{Mode Structure \label{sec:mode_structure}}

\begin{figure*}[t!]
\centering
\includegraphics[width=.35\linewidth]{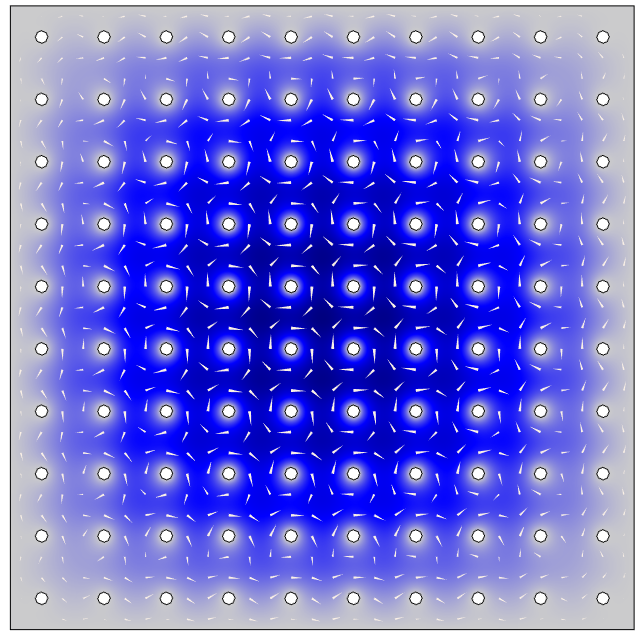}
\label{fig:tm110}
\includegraphics[width=.35\linewidth]{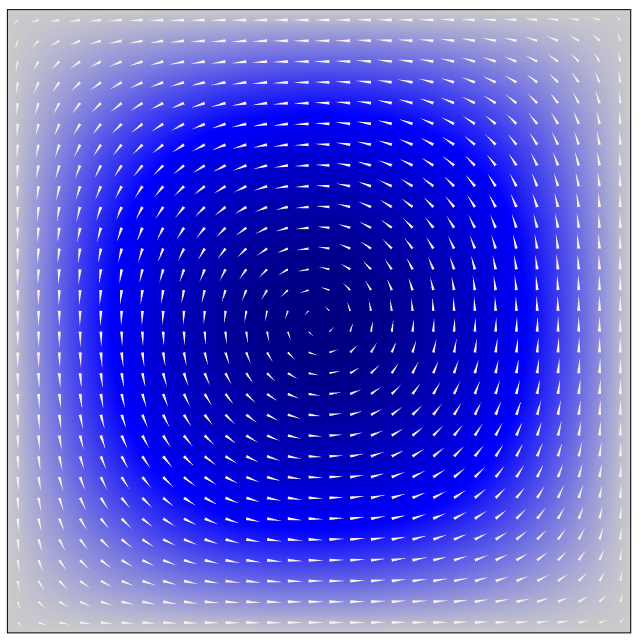}
\label{fig:tm110h}
\includegraphics[width=0.073\linewidth]{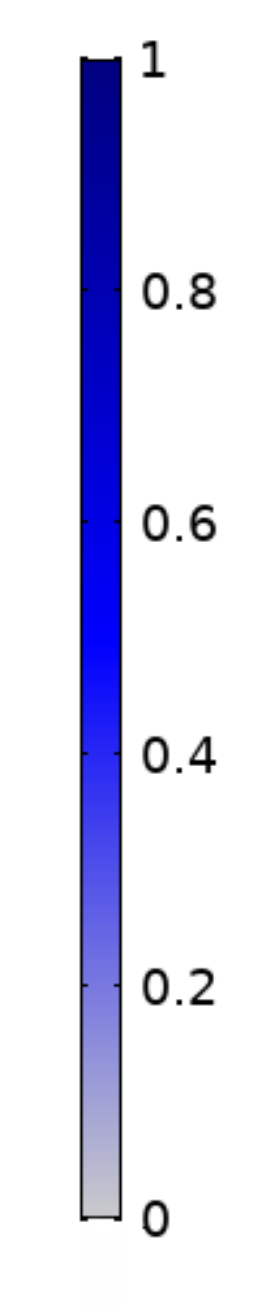}
\hspace{10pt}
    \caption{The TM$_{110}$ mode structure in the $xy$ cross-section of a wire metamaterial loaded cavity calculated in COMSOL Multiphysics. The amplitude of the $E$-field scaled via its maximum value ($E_{\rm max}$), $E/E_{\rm max}$ are shown via the colour map (blue). The log-scaled in-plane components of the $B$-field is shown via the tan arrows. The direction (amplitude) of the $B$-field corresponds to the direction (magnitude) of the arrows.  {\em Left:} A full simulation with metal wires. The wire metamaterial consists of a $10\times 10$ array of circular wires with radius $r=1$\,mm and spacing $a=1$\,cm inside a metal cavity. {\em Right:} A simulation with an effective medium using the analytic formula in Ref.~\cite{Belov2002} using the same parameters, except replacing the wires of the numerical simulation with radius 1\,mm. }
    \label{fig:xy_tm110}
\end{figure*}

\begin{figure}[h]
\centering
\includegraphics[width=1\linewidth]{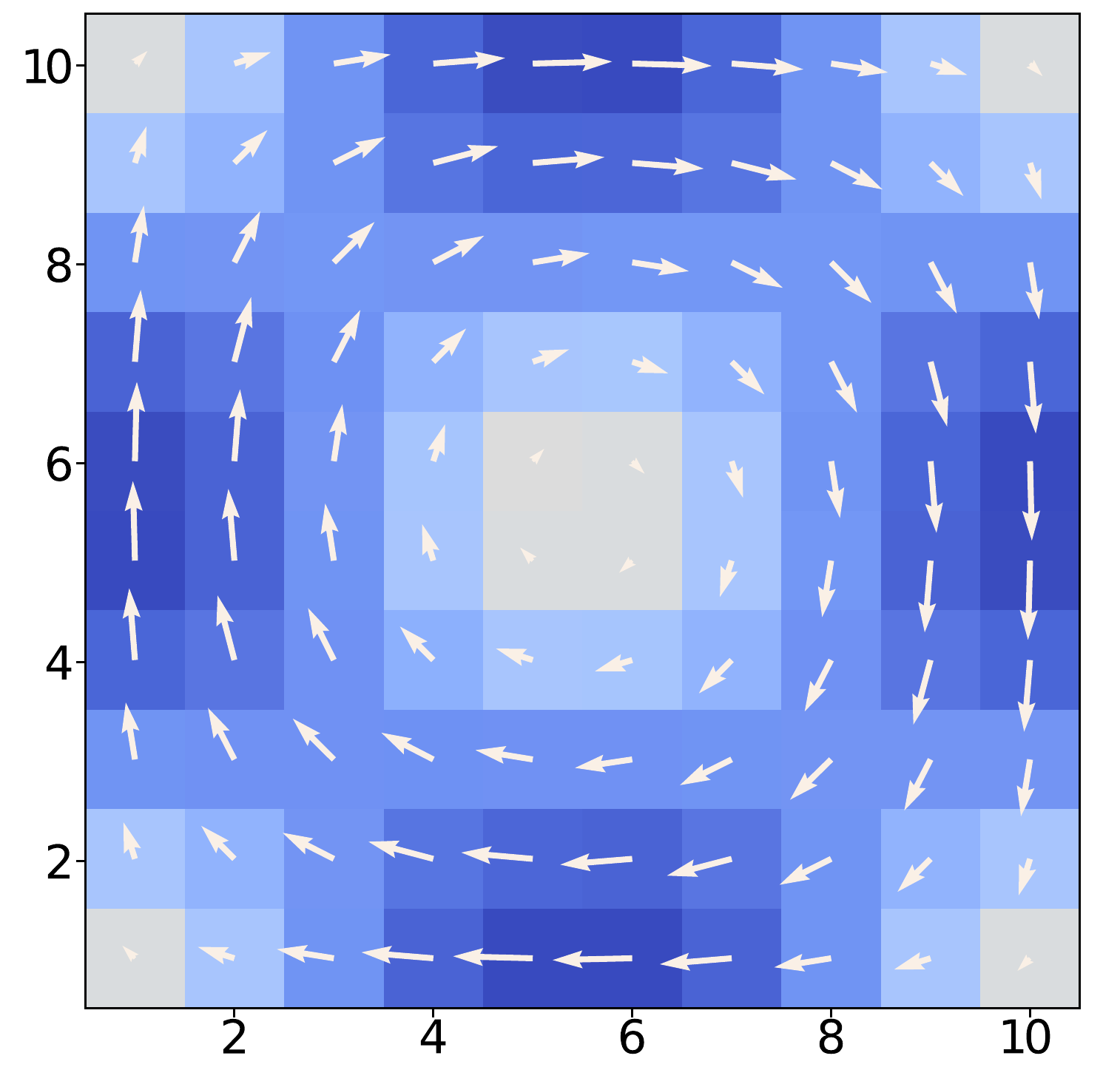}
    \caption{The magnetic field of a cavity with a $10\times 10$ array of wires averaged across each individual unit cell. The arrow direction show the direction of the $B$-field. The magnitude of the $B$-field is shown with the magnitude of the arrows as well as the colormesh plot. One can see that once the local contribution from the wires is averaged out the structure is very similar to the homogeneous case (as shown in the right panel of Fig.\,\ref{fig:xy_tm110}).}
    \label{fig:avg_uc}
\end{figure}

As the mode structure will determine the coupling of the axion to the haloscope, it is crucial that we understand it fully. 
To compare the homogeneous effective medium described in Sec.~\ref{general} with a full wave simulation we show the results of Ref.~\cite{Balafendiev:2022wua} in Fig.~\ref{fig:xy_tm110}. We have plotted the $E$ and $B$-field distributions of the TM$_{110}$ mode. %As previously noted in Ref.~\cite{Balafendiev:2022wua},
We can see that for the $E$-field the overall mode structure of the TM$_{110}$ mode is not affected by the presence of the individual wires. While locally the $E$-field decreases in the immediate proximity of the wires, both cases follow the distribution described by Eqn.~(\ref{eq:TM110}) for overall behaviour. However, with regard to the $B$-fields, there is a significant difference that arises with the local magnetic fields of the wire currents. As a result, the strongest $B$-fields are found not near the walls, as is the case for the homogeneous resonator, but near the center, where the current in the wires is the strongest. Averaging over individual unit cells reveals that the strong local $B$-fields around the wires exist over a weaker background $B$-field (Fig.~\ref{fig:avg_uc}) identical to that of the TM$_{110}$ mode for a homogeneous medium in a cavity. Thus for the coupling to the axion (which depends on the $E$-field) we can be confident that it will be largely captured in an effective medium approach.

Calculating the form factor $\cal C$ gives 0.60 for the full wave simulation, in comparison to 0.66 for the TM$_{110}$ mode of a homogeneous medium, meaning that there is a minimal decrease in signal strength due to the $E$-field vanishing near the wires. While the exact values must always be used for analysing the data, this means that for general use and projecting the discovery potential of an experiment the homogeneous medium approximation will give very accurate results.

The mode structure is also critical to designing an antenna system with an optimal coupling to the system.
Broadly speaking there are two main ways to read out the mode: sampling the $E$-field using a probe or coupled aperture, or sampling the $B$-field using a small magnetic loop. While the $E$-field structure is largely unchanged except immediately around the wires, the maxima of $B$-field occur in a very different location.
Thus, the option of coupling to the cavity near the side using a magnetic probe on the surface seems less compelling, since it will depend strongly on local factors. In addition, an electric probe positioned at one of the walls at which the wires are terminated has the advantage of avoiding coupling to the numerous TE modes of the same cavity. In summary, our simulations indicate that the effective medium approach allows for a good prediction for the overall properties of a wire metamaterial-filled cavity, but that the presence of wires leads to local distortions which will affect the coupling of an antenna.  Understanding the nature of those distortions provides essential guidance in choosing the nature and positioning of readout antennas.

\subsection{Quality Factor \label{sec:quality_factor}}

As Eqn.~\eqref{eq:Q2} was derived under the assumption that the wires were thin (i.e., one dimensional), there may be modifications when applied to relatively thick wires. While Eqn.~\eqref{eq:Q2} was tested numerically in Ref.~\cite{Balafendiev:2022wua}, the simulations used therein for the comparison used a square cross-section wires for easier meshing. %While the numerical and experimental results in Ref.~\cite{Balafendiev:2022wua} were in good agreement, one must be careful to check the behaviour for large, round wires. 
However, the detailed geometry can in fact modify the quality factor. To calculate the quality factor more accurately we simulated wires with a round, rather than square, cross-section.

\begin{figure}[h]
\centering
\includegraphics[width=1\linewidth]{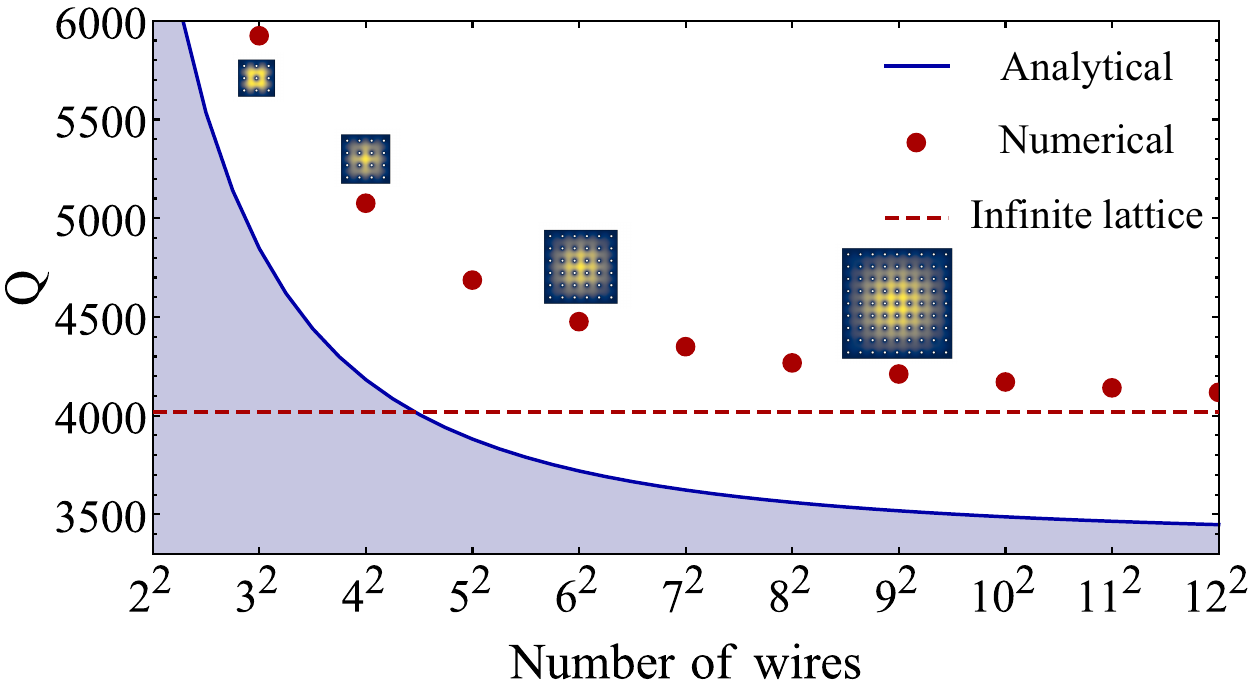}
\label{fig:q_new}
    \caption{Unloaded quality factor $Q$ as a function of the number of wires $N^2$ in a WM consisting of radius $r=1$\,mm wires with a period $a=10$\,mm. Analytic results (blue curve) obtained assuming homogenised medium are compared to the numerical results from the COMSOL eigenmode solver (red points). Quality factor of the infinite wire medium similarly obtained via COMSOL is plotted as the dashed red line. The $xy$ cross-sections of the fundamental mode are shown as insets. The quality factor approaches that of an infinite lattice as the number of unit cells and the size of the cavity increase.} 
\end{figure}

In Fig.~\ref{fig:q_new}
we show how the quality factor changes as the size of the resonator is increase, asymptoting towards the value of an infinite 2D structure (simulated in COMSOL). Here, we see that the quality factor is actually higher by 30\% when compared to the analytic estimates. This increase would lead to a significant enhancement of the experiment, particularly at high frequencies where requiring that the system is single-moded may limit the volume of the experiment. One possible explanation is that the sharp corners in the square wires leads to concentrations of current and so larger surface losses. While this increase remains to be confirmed experimentally with high quality prototypes, these simulations indicate that Fig.~\ref{fig:radius} is actually conservative and that noticeably higher quality factors can be reached.

\subsection{Tuning}
\label{sec:tuning}
As the axion mass is unknown, a key feature of any resonant experiment must be the ability to tune and search a range of frequencies.
The initial proposal in Refs.~\cite{Gelmini:2020kcu,Wooten:2022vpj} for tuning was to adopt a rectangular wire array, moving planes of wires relative to each other in a single direction. While this gave a significant range of tuning, it had the disadvantage of changing the volume of the wire array. Not only would this decrease the possible signal power, but also create large air gaps if surrounded by a cavity, potentially changing the mode structure, for example allowing modes to form in the air gaps.

\begin{figure}[h]
\centering
\includegraphics[width=1\linewidth]{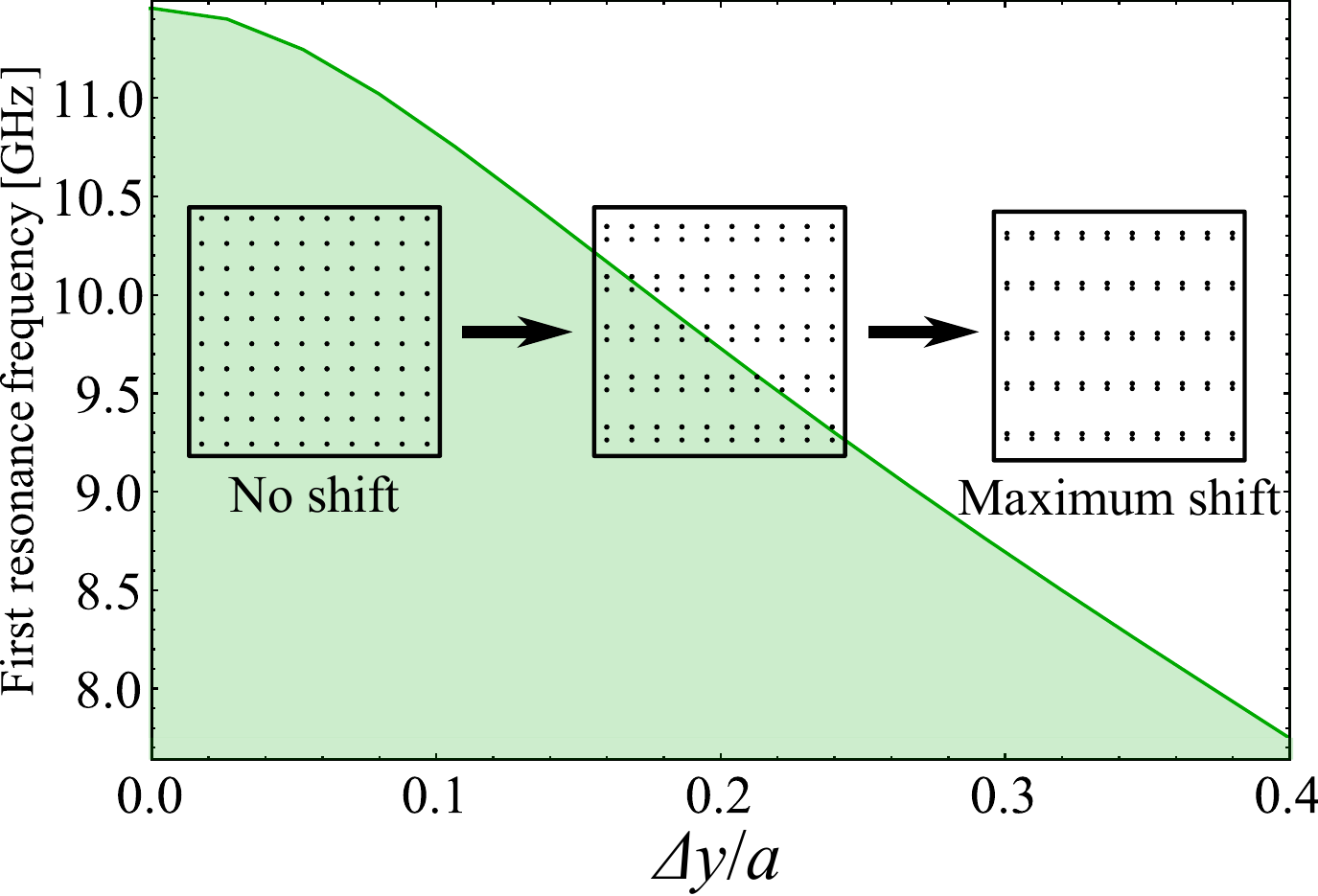}
\caption{Tuning of the resonance frequency by moving pairs of wires together. Here $\Delta{y}$ is the distance by which each row of wires is moved and $a$ is the initial period between wires. The change of frequency with the maximum offset of wires constitutes about 30\% of the initial value. This tuning would easily facilitate a broad scan in frequency space. Insets show the relative positions of the wires graphically for three different stages of tuning. Here the radius of wires $r$ is $1$\,mm and the period $a=10$\,mm}.
\label{fig:tune}
\end{figure}

\begin{figure}[h]
\centering
\includegraphics[width=1\linewidth]{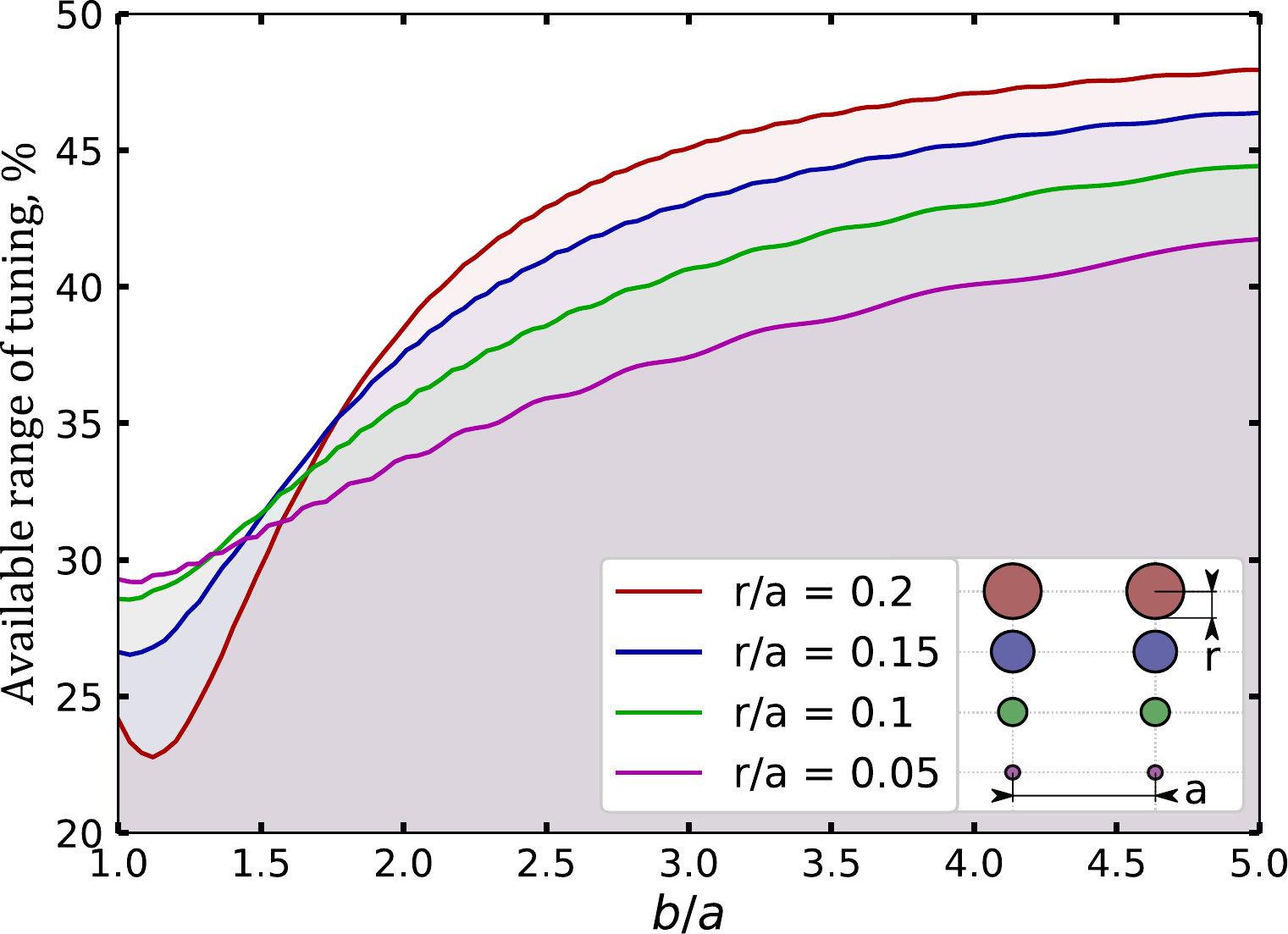}
    \caption{Maximum available tuning range, $(f_0 - f)/f_0$ where $f_0$ corresponds to the untuned case vs the initial lattice aspect ratio $b/a$. Larger aspect ratios (larger differences in wire periods in the $x$ and $y$ directions) allow for more tuning. We show a variety of wire radius, showing $r/a=0.05,0.1,0.15,0.2$ in purple, green, blue and red, respectively. } \label{fig:tune_range}
\end{figure}

In order to avoid these issues, we propose a new method to change the mutual inductance. %\Alex{Where has it been shown? Or is it being shown here?}
One can split the lattice into two rectangular sublattices consisting of odd and even rows of wires, allowing the  the two to be brought together, lowering the plasma frequency of the system. In other words, rather than moving planes of wires relative to each other, we move pairs of wires. Preliminary simulations in COMSOL eigenmode solver show (Fig.~\ref{fig:tune}) that  mechanical tuning in this fashion can allow for up to a 30\% change in the frequency. The exact extent of the tuning is determined by the aspect ratio of the sublattices and the radius of the wires in relation to the period, with larger aspect ratios and wire radii potentially allowing for more tuning, as shown in Fig.~\ref{fig:tune_range}. Crucially, there is a minimal change to the active volume of the wire medium, as shown in the insets of Fig.~\ref{fig:tune}, and thus minimal loss of signal power. Such a procedure should also protect against alterations to the mode structure and new modes forming, helping to avoid unfortunate gaps in tuneability. As there is only a single degree of freedom it is possible to design a one dimensional tuning system, reducing the potential for mechanical or tuning errors. 

\section{Experiment Prototypes\label{sec:experiment}}
Experimental prototypes are of critical importance to validate the analytical and numerical results. Moreover, such prototypes would reveal practical challenges of operating a plasma haloscope, which may be hidden in the simplified models.
So far, studies have been done in both free space~\cite{Pendry:1998,Belov:2003,doi:10.1063/1.1513663, Wooten:2022vpj} and inside a cavity~\cite{Balafendiev:2022wua}. Here we review the results of the prototypes pursued under the ALPHA program, Refs.~\cite{Wooten:2022vpj,Balafendiev:2022wua}.
\subsection{Measurements in Free Space}
The cleanest environment to explore the pure plasma properties of a metamaterial is in free-space, removing potential complications due to cavity walls and cavity modes. First studies showed that wire media indeed demonstrated a plasma behaviour~\cite{Pendry:1998}, albeit exhibiting spatial dispersion~\cite{Belov:2003}. Here we summarise the results in free space, focusing mainly on the recent Ref.~\cite{Wooten:2022vpj}.

The simplest way to gauge the electromagnetic properties of a wire metamaterial is to perform a reflection/transmission measurement. Such a measurement allows us to see the transition between dielectric and metallic behaviour, as well as study the effective properties of the system such as the quality factor. 

 Following Ref.~\cite{Wooten:2022vpj}, for normal incidence the transmission $S_{21}$ through a slab with complex dielectric constant can be expressed as 

\begin{equation}
S_{21} = \frac{(1-\rho_{0}^{2})e^{-(\alpha +j\beta)d}}{1-\rho_{0}^{2}e^{-(2\alpha+j\beta)d}}\,,\label{eq:s21}
\end{equation}
where $\rho_{0} =(1-\sqrt{\epsilon})/(1+\sqrt{\epsilon})$, \emph{d} is the width of the slab, and $\alpha$ and $\beta$ are the attenuation coefficient and wave numbers of the propagating wave in the medium, respectively. The $\alpha$ and $\beta$ are given by
\begin{subequations}
\label{eq:whole}
\begin{equation}
\alpha(\nu) = \frac{2\pi\nu}{c}\sqrt{\frac{\epsilon'}{2}\left(\sqrt{1+\tan^{2}\delta} -1\right )}\,,
\end{equation}
and
\begin{equation}
\beta(\nu) = \frac{2\pi\nu}{c}\sqrt{\frac{\epsilon'}{2}\left (\sqrt{1+\tan^{2}\delta} +1\right )}\,,
\end{equation}
\end{subequations}
where $\tan \delta=\epsilon''/\epsilon' $  is the loss tangent of the material.  By comparing measurements of a given wire metamaterial's $S_{21}$ data with a fit using Eqn.~\eqref{eq:s21} one can then find the plasma frequency and quality factor, with the most accurate results coming from the transition region between metallic and dielectric behaviour (i.e., at the plasma frequency).

\subsubsection*{Experimental setup and measurements}
\begin{figure*}[ht!]
\centering
\includegraphics[width=180mm]{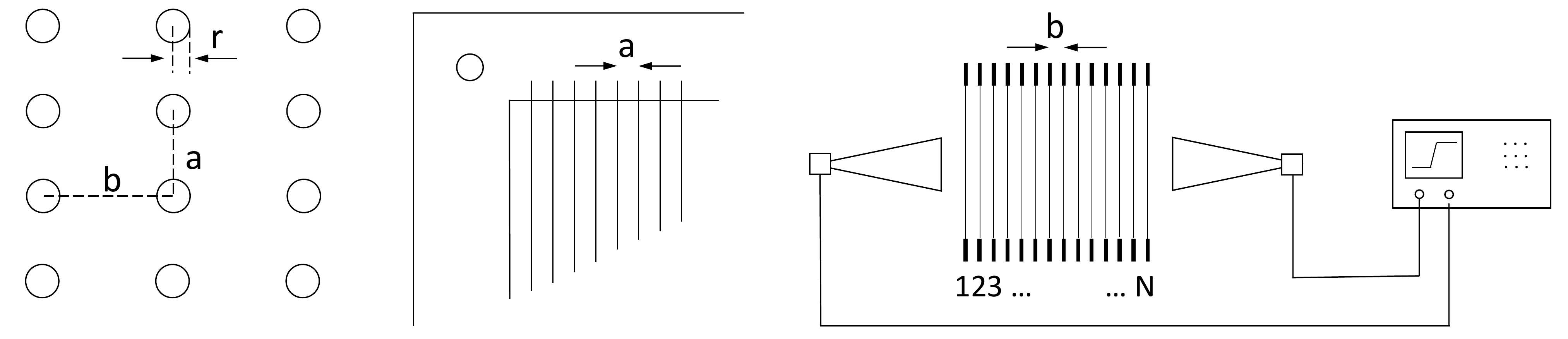}
\caption{\label{fig:berk1} {\em Left:}  A rectangular wire array, with $r$, \emph{a} and \emph{b} designating the radius and the spacing of the wires in each direction, respectively. {\em Middle:}  Detail of the construction of a single wire plane.  The wires are supported by a metal frame but they are not in electrical contact with it.  {\em Right:}  The geometry for measuring the transmission $S_{21}$ as a function of number of wire planes, \emph{N}. Horns are used to transmit and receive microwave signals, which is then analysed by a VNA. Adapted with permission from~\cite{Wooten:2022vpj}.}
\end{figure*}

The free-space experiment consisted of 40 wire planes consisting of square metal frames of 203 (254)\,mm inner (outer) edge length, 1.5\,mm thickness, and strung with 50\,$\mu$m diameter gold-on-tungsten wire~\cite{Luma} with a wire spacing of $\emph{a} = 5.88$\,mm. The $S_{21}$ measurements were carried out with a vector network analyzer (VNA)~\cite{keysight} and matched waveguide horn antennas~\cite{Pasternak}. This setup is illustrated in Fig.~\ref{fig:berk1}. 

There are several important questions that must be first asked when looking at wire metamaterials. The first is how many wires are required for the system to behave as an effective medium. This was first explored in Ref.~\cite{doi:10.1063/1.1513663}, and further studied in Ref.~\cite{Wooten:2022vpj}. It is also important to know how the quality factor changes as the number of frames increases. This is because in the absence of a cavity energy can exit the system both by radiation and restive losses, though for very low numbers of wires the effective medium description may be unreliable. In the Berkeley experiment (Ref.~\cite{Wooten:2022vpj}) successive wire frames were added, up to a total of 40 frames.

\begin{figure}[ht!]
\centering
\includegraphics[width=.5\textwidth, trim = 6cm 3cm 5cm 3cm, clip, scale=0.5]{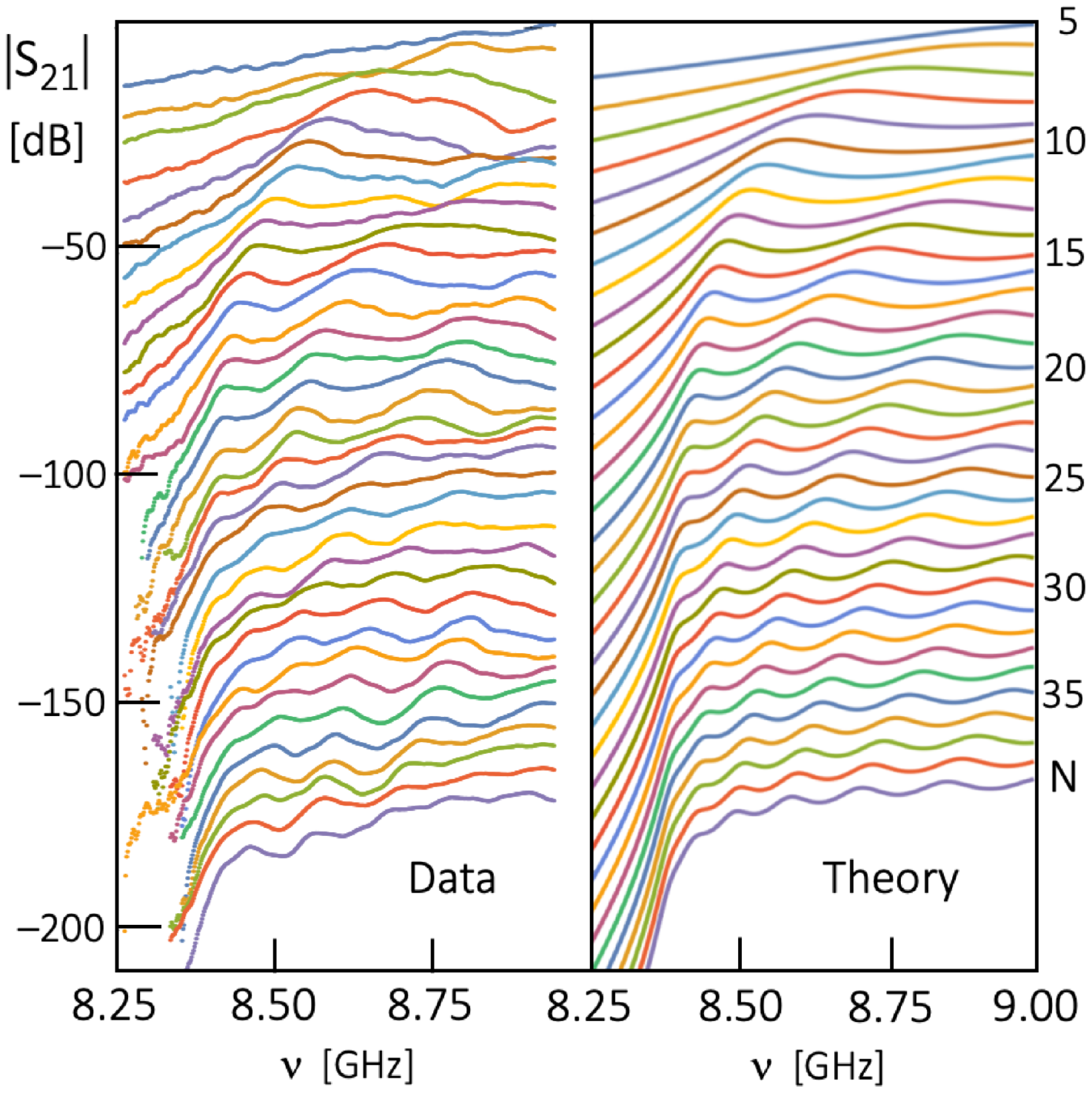}
\caption{\label{fig:berk2} Ensemble of measured and calculated $S_{21}$ for all \emph{N}; spectra are sequentially displaced by -5 dB for clarity.  The calculated spectra are all for $\nu_{p}$ = 8.35~GHz,  $\mathit{\Gamma}$ = 0.04~GHz, and the array width \emph{d} = $N\cdot b$. Taken with permission from Ref.~\cite{Wooten:2022vpj}.}
\end{figure}

The wire frames were loaded so that the wire media was formed from a  rectangular array with (\emph{a},\emph{b}) = (5.88, 8.00) mm (Fig.~\ref{fig:berk1}) The measurements results are shown in Fig.~\ref{fig:berk2}, contrasted with the fitted estimates using Eqn.~\eqref{eq:s21}. We can see that the theory is in good qualitative agreement with all cases.

%Least squares fitting of the $S_{21}$ of a frequency-dependent complex dielectric yields an effective width of the full array equal to its physical width at the few percent level (Figure 4a).

\begin{figure}[ht!]
\centering
\includegraphics[width=.55\textwidth, trim = 8.45cm 3.5cm 6.5cm 3cm, clip, scale=1]{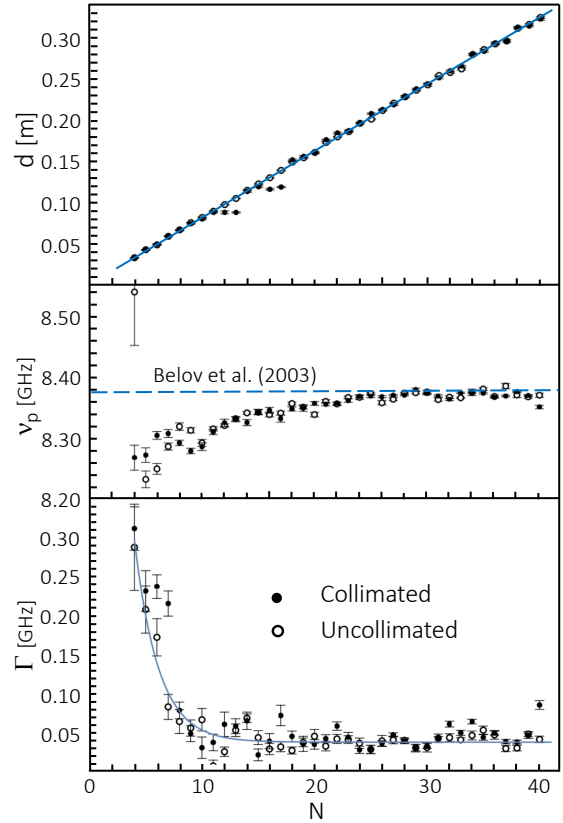}
\caption{\label{fig:berk3}Metamaterial parameters derived from the measured $S_{21}$ as a function of the number of planes, \emph{N}. {\em Top:} The effective width of the wire array metamaterial, \emph{d}.  {\em Middle:}  The plasma frequency $\nu_{p}$. As can be seen, the semianalytic theory of Ref.~\cite{Belov:2003} shown as a blue line is in excellent agreement with the asymptotic plasma frequency, within 0.1\% for $N > 25$. {\em Bottom:} The loss term, $\mathit{\Gamma}$. Adapted with permission from Ref.~\cite{Wooten:2022vpj}.  }
\end{figure}

To see the quantitative agreement, we show the parameters as a function of the number of planes $N=1-40$ Fig.~\ref{fig:berk3}. Interestingly, the effective size of the medium is equivalent to $N+1$ layers: in other words, the effective media extends \emph{b}/2 on either side of the array. As noted in Ref.~\cite{Wooten:2022vpj} the plasma frequency appears well defined after only approximately 5 layers, and exhibits little change after 10 layers with excellent agreement with the theoretical prediction (within 0.1\% for large $N$). Thus even relatively small arrays behave as a well defined effective medium, similarly the
behaviour shown in a cavity~\cite{Balafendiev:2022wua}.

Another crucial factor is the quality factor or loss rate of the system. Unlike in a cavity, the system can also experience losses where power is radiated from the system. Further, for a very small number of wire planes there may be some surface effects, or the wires may not yet behave as an effective medium. As the wires used were relatively thin, one would expect resistive losses to be more significant, with $Q=260$ at room temperature. In the bottom panel of Fig.~\ref{fig:berk3} we can see that the loss rate quickly decreases as wire frames are added, most likely representing the transition to an effective medium. This then asymptotes to a $\Gamma$ corresponding to $Q=220$, only $15\%$ lower than the predicted value. As discussed in Section~\ref{sec:losses} these losses can be improved by up to two orders of magnitude by using thicker wires in a cryogenic environment. 

We can thus see that the simple analytic model describes accurately the plasma properties of wire metamaterials, allowing us to extrapolate to plasma haloscopes which operate outside of a cavity, as may be desirable to avoid TE modes.

\subsection{Measurements in Brass Cavity}
In order to avoid radiative losses and maximise the quality factor, it may be advantageous to place the metamaterial inside a conductive cavity. However, this also adds a new element of complexity to the setup, the interaction of the wire material with the cavity walls. To verify that the analytic and numerical formalisms for plasma halsocopes holds in cavities,  Ref.~\cite{Balafendiev:2022wua} constructed and measured a brass prototype in ITMO University.

\begin{figure}[h]
\centering
\includegraphics[width=.49\linewidth]{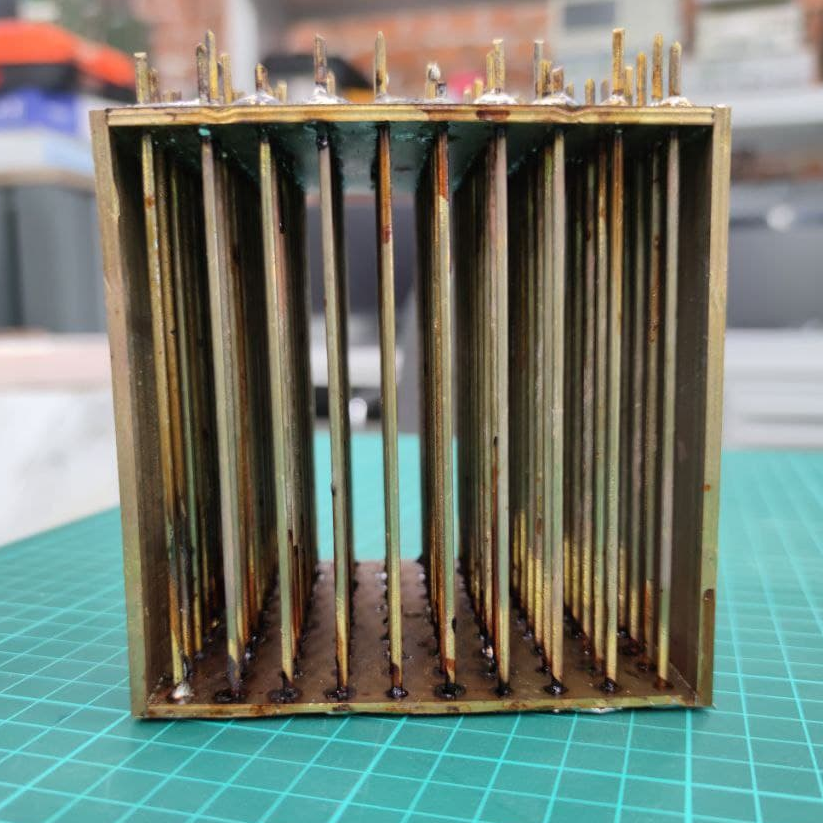}
\label{fig:itmo_exp_1}
\includegraphics[width=.49\linewidth]{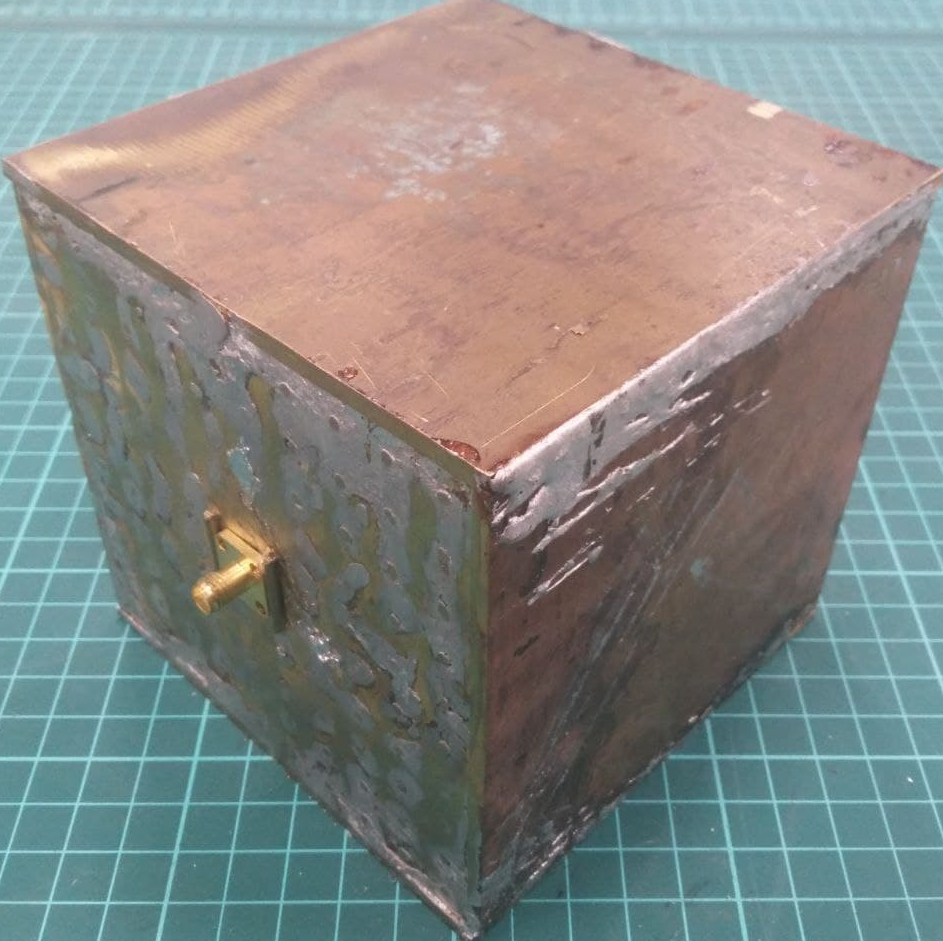}
\label{fig:itmo_exp_2}
    \caption{Photos of the experimental prototype cavity loaded with wire metamaterial in Ref.~\cite{Balafendiev:2022wua}, reproduced with permission. The plasma haloscope consists of a 10$\times$10 array of radius $1$\,mm circular cross-section brass wires placed with a period $a=1$\,cm. The wires are inserted into the holes in the brass walls and soldered to ensure an electric connection. Two 8-mm long SMA connectors acting as monopole antenna probes are inserted at the centers of the walls to which the wires are connected. {\em Left:} The wires inside the plasma haloscope before final assembly. {\em Right:} The finished resonator fully enclosed in a cavity. Taken with permission from Ref.~\cite{Balafendiev:2022wua}. }
    \label{fig:itmo_exp}
\end{figure}

In the setup of Ref.~\cite{Balafendiev:2022wua} (shown in Fig.~\ref{fig:itmo_exp}), yellow brass wires (65\% copper, 35\% zinc) of $r=1$\,mm  were placed and welded within a $10\times 10\times 10$\,cm brass cube with a period $a=1$\,cm. The lateral walls were spaced $a/2=0.5$\,cm away from the wire metamaterial sample. This wall spacing matched the effective size of the medium to the cavity space. The measurements were performed with two SMA (sub-miniature A-version) ports acting as monopole antennas, allowing for the frequency response and quality factors to be measured with an Agilent E8362C VNA. 

\begin{figure}[t]
\centering
    \includegraphics[width=\linewidth]{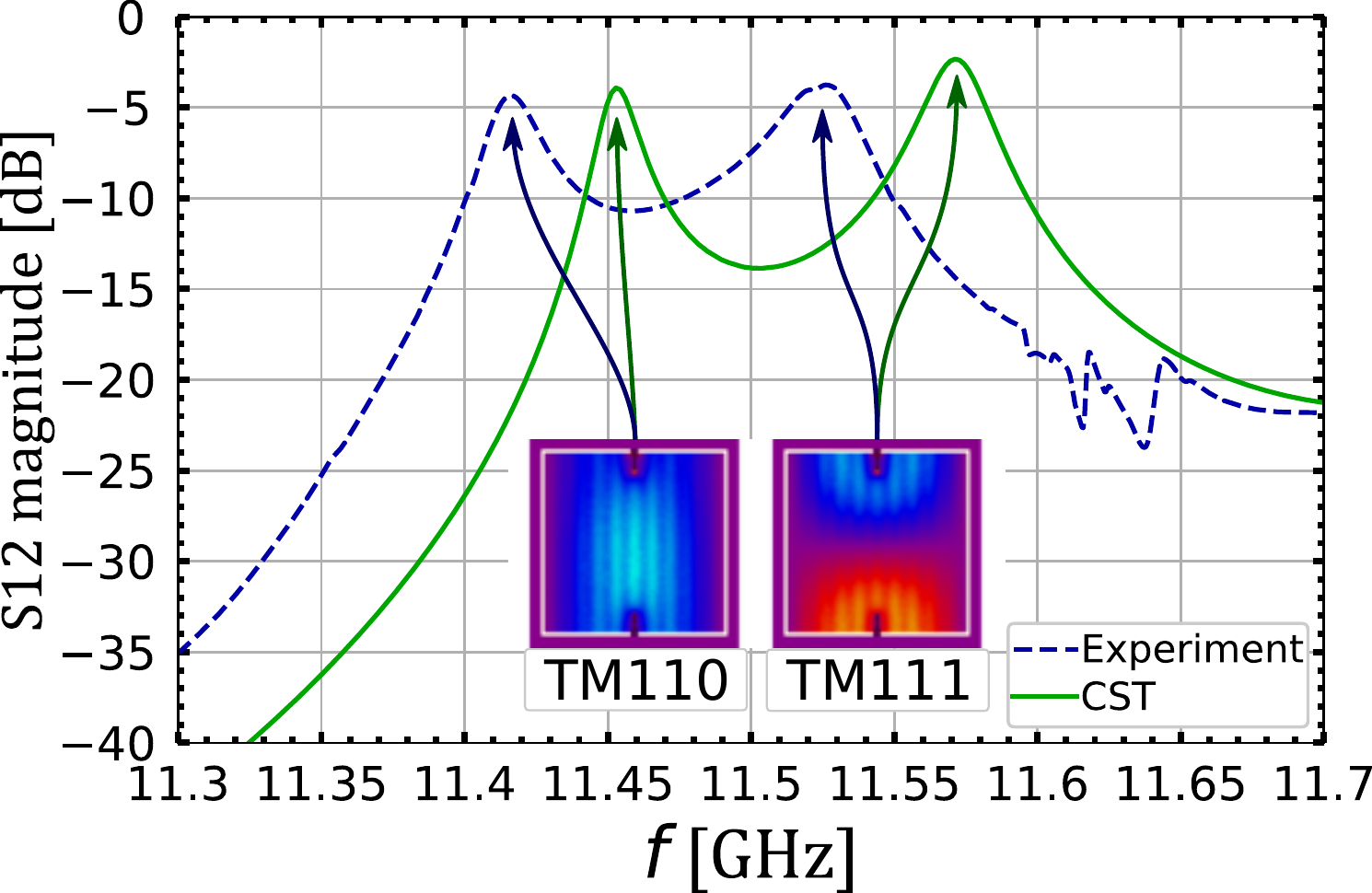}
    \caption{S12 (transmission) parameters for cavity with a $10\times 10$ wire array with spacing $a=1$\,cm inside of it, taken from Ref.~\cite{Balafendiev:2022wua}. Ref.~\cite{Balafendiev:2022wua} performed both a CST simulation (green line) and the measured values of the prototype experiment (blue dashed line) as shown in Fig~\ref{fig:itmo_exp}. The CST simulation assumes 65\% brass wires with a circular cross-section of radius 1\,mm, whereas the experiment uses the same geometry wires but with a measured DC conductivity of $1.51\times 10^7$\,S/m (slightly lower than 65\% brass). The lowest order TM modes are depicted with $xz$ cross-sections in the insets. }
    \label{fig:s12}
\end{figure}

We reproduce the measurements of Ref.~\cite{Balafendiev:2022wua} here in Fig.~\ref{fig:s12}. They found an agreement in the resonance frequency within 1\% with numerical simulations in CST. The primary difference occurred in the unloaded quality factor,  which was 40\% lower
 than the simulations. This agreement is very good considering that the primary focus was on validating the mode structures of the numerical simulations, and so maximising the quality factor was not a priority. There was a similar disagreement in the coupling coefficients, which may have been due to the non-ideal electrical connection of the port which could have leaked energy, resulting in a lower coupling. These parameters are summarized in Table~\ref{table:itmo-proto}.

The extra losses in the experimental setup may have been due to several reasons, most of which arose due to it being a simple proof of principle prototype. It is possible that the outer layer of the wires oxidized, changing the AC conductivity but leaving the DC conductivity largely unchanged due to the skin effect. The surface roughness of the wires may also have been insufficient, or excess solder could have been present. It is also possible that there were gaps with the readout antenna, or that wires were misaligned. Future prototypes will correct these issues.

  \begin{table}[t]
\centering
\begin{tabular}{@{}lcc@{}}

%TC:ignore
\textbf{}                                 & \multicolumn{1}{l}{\textbf{Experimental}} & \multicolumn{1}{l}{\textbf{Numerical}} \\ \midrule
\textbf{Frequency {[}GHz{]}}              & 11.420                             & 11.453                               \\
\textbf{Bandwidth {[}GHz{]}}              & 0.022                              & 0.016                                 \\
\textbf{Loaded Q}                         & 509                                & 735                                \\
\textbf{Coupling coefficient}             & 1.34                               & 1.82                                \\
\textbf{Unloaded Q}                       & 1194                               & 2074                               \\ \bottomrule
%TC:endignore
\end{tabular}

\caption{Comparison between numerical simulations in CST and measurements of the experimental prototype of Ref.~\cite{Balafendiev:2022wua} for the key parameters of the TM110 mode. The plasma haloscope consists of a $10\times 10$ wire array with spacing $a=1$\,cm inside a metal cavity. The CST simulation assumes 65\% circular brass wires with radius $1$\,mm, whereas the experiment uses wires with the same radius but has a measured DC conductivity of $1.51\times 10^7$\,S/m (slightly lower than 65\% brass).}
\label{table:itmo-proto}
\end{table}

\subsection{Measurements in Copper Cavity}
\begin{figure}[ht!]
\centering
\includegraphics[width=.49\linewidth]{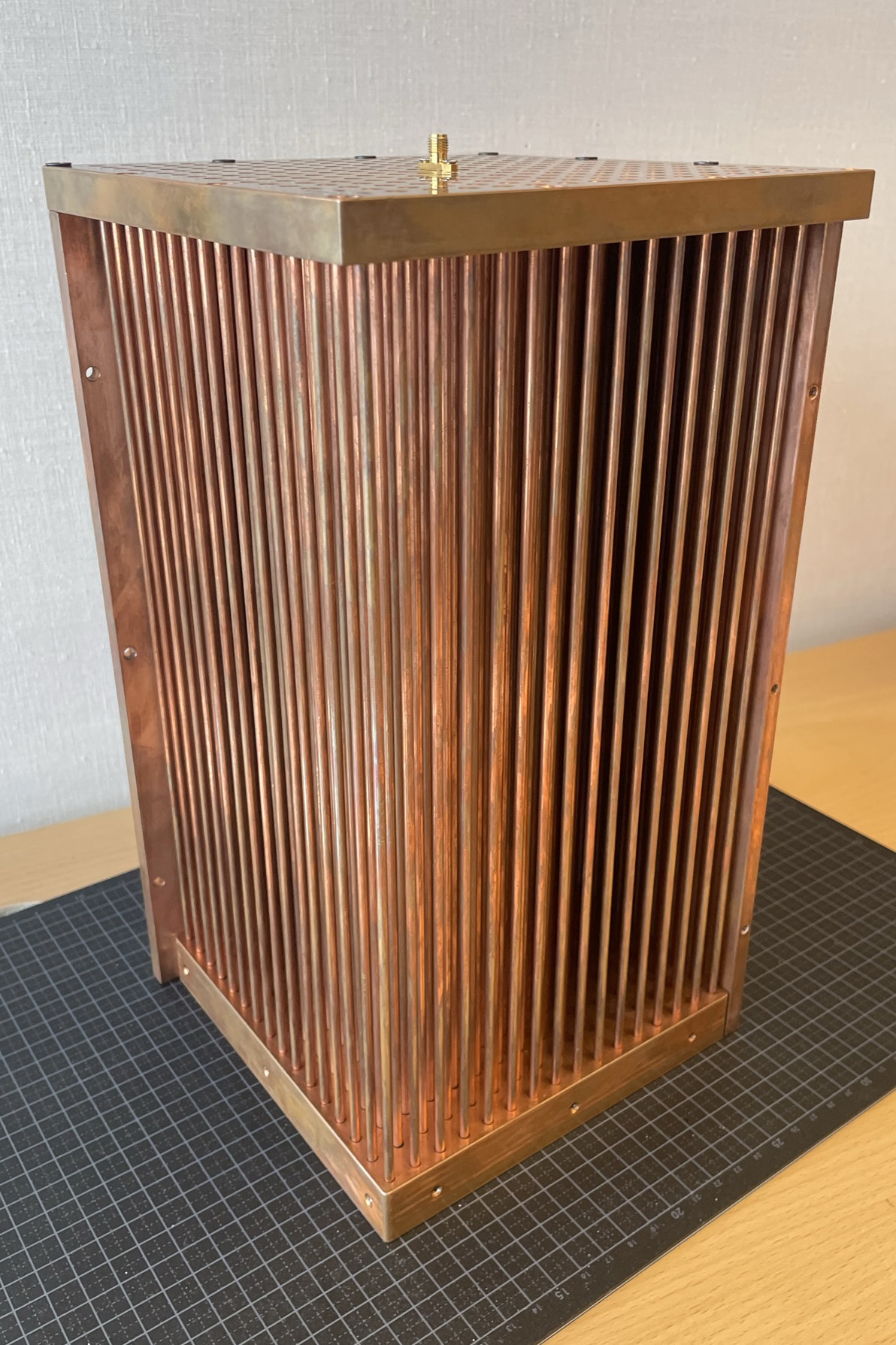}
\label{fig:su1}
\includegraphics[width=.49\linewidth]{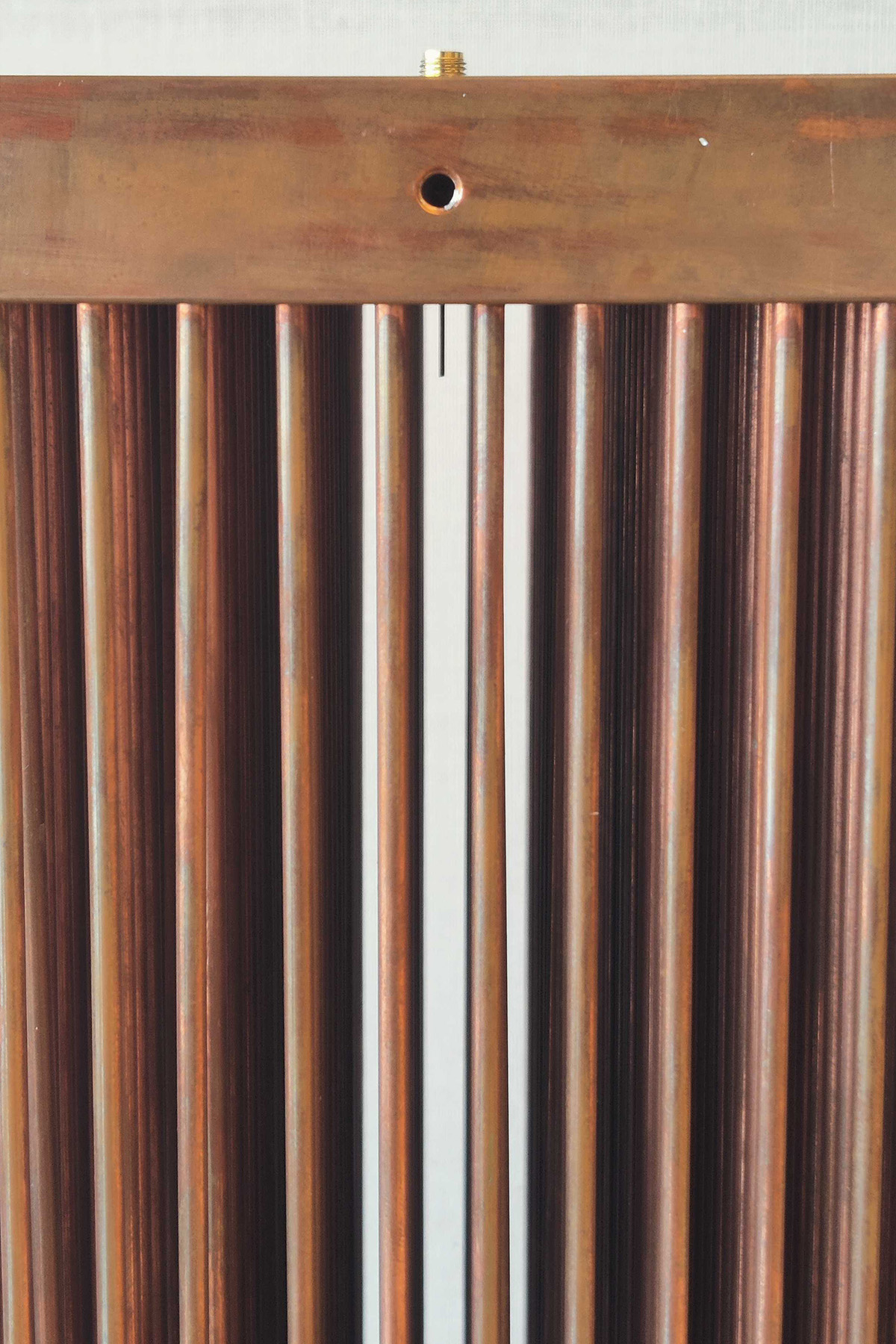}
\label{fig:su2}
\caption{ Left: Photo of the $16 \times 16$ wire copper prototype with two of the four walls removed. Right: Closeup of one of the two antenna probes showing its extension into the cavity.}
\label{fig:su}
\end{figure}

In order to explore some technical aspects of the cavity design, a copper prototype with a $16 \times 16$ wire array has been built at Stockholm University (see Figure \ref{fig:su}). The cavity, which was built using C110 copper, has rectangular inner dimensions of $160.2 \times 160.2 \times 290.5$\,mm. The copper wires are made from 3.18-mm outer diameter tubing which was then threaded to accommodate mechanical and electrical connections to the top and bottom plates of the cavity. Unlike the brass prototype, the four walls of this cavity can be detached from the top and bottom plates. This allows us to more easily make modifications to the system. Circular holes drilled into the top and bottom plates accommodate the insertion of simple antenna probes for RF measurements. To support the array assembly, we chose to make the top and bottom plates of different thickness. This also means that the top and bottom antenna probes must be threaded through slightly different geometries as they pass into the cavity. This has implications for RF modelling due to the reduced symmetry of the system. This copper prototype has been characterized using a ZNA26 from Rohde \& Schwarz. 

Because of the seven times volume increase over the brass prototype, it is more challenging to produce 3D simulations that enable fair comparison with measurement results than in the case of the smaller brass prototype. The 3D simulations that have been produced so far predict an unloaded quality factor for the TM110 mode of approximately 3000. We expect this number to rise with higher-fidelity simulations, especially as meshing resolution is improved. Ongoing S-parameter measurements show promising results, and will be reported in future work.

\section{Superconducting Wires}
\label{sec:superconductors}
While the performance of normal conducting metamaterials is extremely promising, it is important to consider possible ways to upgrade an experiment. One intriguing possibility would be to replace the conducting wires with superconducting wires.
The application of superconducting materials for a different class of axion experiments have been explored in the conventional resonant cavities~\cite{9699394, Posen_2022, PhysRevApplied.17.L061005}. In the past, it was believed that a $Q$ higher than the intrinsic axion $Q\sim 10^6$, estimated from  the Maxwell-Boltzmann distribution of the galaxy halo, may be of no use because such a high-Q cavity would filter out the axion signal. Recent studies revealed that the signal power is enhanced even with $Q>10^7$ and thereby scanning rate is increased with higher $Q$~\cite{Kim_2020}. The best $Q$ of superconducting cavities under a strong static magnetic field was given by high-T$_c$ superconducting tapes~\cite{PhysRevApplied.17.L061005}. The superconducting wire metamaterial is therefore a promising option toward further improvement of the sensitivity of the plasma haloscope.

Superconducting wires have a number of potential advantages for use in a plasma haloscope.  First, superconductors have a frequency-independent screening length, called the magnetic penetration depth $\lambda$, which is on the scale of ${\cal O}(10)-{\cal O}(100)$\,nm at low temperatures~\cite{Tinkham2004,Anlage92,Anlage2021}.  This is in contrast to both the ordinary and anomalous skin effect in normal metals, where the screening length, and associated skin-effect losses, are frequency dependent.  The frequency-independence of the superconducting screening length is guaranteed for frequencies smaller than the gap frequency $\Delta/h$, (where $\Delta$ is the superconducting energy gap and $h$ is Planck's constant) typically at ${\cal O}(100)$\,GHz.  

Secondly the superconducting surface resistance $R_{\rm s}$ can in principle be made arbitrarily small by decreasing the number of quasiparticles (excitations out of the ground state), whose number density $n_{QP}$ scales as $n_{QP}(T) \sim e^{-\Delta/k_{\rm B} T} \rightarrow 0$ for $T \rightarrow 0$~\cite{Anlage2021,PhysRev.111.412}.  The first demonstration of plasma response of superconducting Nb wire arrays showed quite low losses \cite{Ricci2005}.  In practice there is an extrinsic residual resistance that interrupts the decrease of losses~\cite{PhysRevB.96.184515}, but this is often not a practical concern except in the most extreme high-$Q$~\cite{PhysRevApplied.13.014024} or quantum-coherent applications of superconductors.

In the case of plasma haloscopes, losses from the walls of the system would likely dominate in this regime. Under such conditions the factors determining the quality factor become quite similar to the case of a conventional cavity. As such, the unloaded quality factor would be approximately two orders of magnitude higher than a WM made of copper wires, with the $Q$ growing linearly with increasing the volume. This scaling is because the energy stored grows as the volume, but the power lost is proportional to the surface area of the cavity walls. However, detailed studies of the superconducting case are necessary to reliably predict quality factors under experimental conditions, and other concerns such as the precision of the wire placement may also influence $Q$.

An interesting possibility of superconductors is their large degree of tunability, enabled by multiple physical mechanisms. As the plasma frequency is determined by the mutual inductance of the wires, any way of modifying this inductance will necessarily tune the device. The kinetic inductance of superconductors is tuneable through variations in the superfluid density with temperature, dc current, and dc magnetic field \cite{MesTed,MasonGould,Anlage89,Ann2010,Clem2012,Kurter2013,Vis2015,Adam2016,Frasca2019,Xu2019,Mash2020,Oripov2021}.  The first demonstrated tunability of superconducting wire dielectric properties was demonstrated using microwave power and dc magnetic field \cite{Ricci2007}.  The Josephson effect enables large tunability of inductance with near-zero added loss \cite{Chris71,Rifkin76,Trep2013,Cast2008,ZhangD2015,Nguyen2019}. In fact, this method of tuning has already been demonstrated for WM~\cite{doi:10.1063/1.5126963} utilizing a dc current.  Granular superconductors take advantage of a natural array of Josephson junctions to create high tunability \cite{PhysRevApplied.5.044004,Grun2018,Maleeva2018,Oripov2021}, and are frequently used as elements in superconducting single-photon detectors~\cite{Day2003}.  Finally, the proximity effect between a superconductor and normal metal can create strongly temperature dependent electromagnetic properties of the composite system, and is well suited for a wire geometry \cite{deGennes,Oda80,Mota82,Pam94,Pam95,Pam96,Pro2001}. These methods are an interesting line of future work for tuning plasma haloscopes non-mechanically, which may be particularly useful at high frequencies. However, it must be verified that tuning the superconductor does not lead to degradation in other key areas, for example quality factors~\cite{Eom2012}.  %All of these tuning mechanisms are discussed in detail in the document entitled ''ALPHA Superconductors Working Group Internal Technical Report."

An important thing to note about superconducting tunability is the fact that most of the demonstrations have been performed on thin films, which are typically prepared on flat dielectric substrates and patterned with photo- or nano-lithography.  The high dielectric constant $\epsilon$ of the substrates enables the wire arrays to be shrunk in each linear dimension by a factor of $\sqrt{\epsilon}$, which can be a factor of 3 to 5. Unfortunately, such a dielectric constant would also screen the axion induced $E$-field, which may lead to a loss in signal. It is possible that multiple compact superconducting haloscope units could be designed to operate at different center frequencies and reside in the same magnet.

An alternative approach is to work in the more traditional cylindrical wire geometry by depositing superconducting thin films on either metallic or dielectric wires and relying on the above tuning mechanisms that work without lithographic patterning. This has the advantage of avoiding possible issues with a dielectric substrate either screening the axion induced $E$-field or potentially altering the mode structure. However, any limitations of mechanical tuning at very high frequencies will also apply to such a set up. 

%Estimate of quality factor for a static (non-tuned) superconducting wire array is highly non-trivial.  The surface resistance caused quasiparticles is approximately proportional to $\omega^2$ for $\omega << \Delta$.

One concern with using superconducting wires is their performance in high magnetic fields.  Thin film superconductors of thickness $t$ experience an enhanced critical field when subjected to a parallel magnetic field when $t<\lambda$~\cite{Tinkham2004}.    A thin wire superconducting loop-gap microwave resonator has been created and shown to maintain a high-Q resonance ($Q_i > 2\times 10^5$) in parallel magnetic fields up to 6 T~\cite{PhysRevApplied.5.044004}. The resonator is made from NbTiN film that is 8~nm thick, patterned into a wire about 100~nm wide, resonating at 3 to 5~GHz at a temperature of 280~mK \cite{PhysRevApplied.5.044004}. This is a proof of principle that superconducting wires can maintain high-Q microwave resonances in multi-Tesla parallel fields.  Superconducting NMR pickup coils in the 100 MHz - 1 GHz range made of thicker films have been shown to operate well in parallel fields of 1-2 T \cite{Hill1997,Brey2006,Ram2013}.  Note that dielectric-loaded compact wire arrays can be kept in regions of magnetic field with minimal radial divergence, thus preserving their low-loss properties. Alternatively, when searching for HP DM a magnetic field is not required, meaning that a HP only plasma haloscope could sidestep the issue of high magnetic fields. Superconducting wires are thus an interesting avenue of future work.

%\Akira{probably the present state-of-the-art is a Korean result with HighTc SCs:https://indico.him.uni-mainz.de/event/109/contributions/834/}
%\Akira{I would also mention Dongok Kim et al JCAP03(2020)066, which showed that we gain scanning rate even if the cavity Q is higher than axion Q of 1e6}
\section{Experiment Design}
\label{sec:design}
Putting the previous considerations together, we can now specify a baseline design for the ALPHA experiment, as well as indicating possible upgrade paths. For axion searches, the most significant infrastructure is, of course, the magnet. For our benchmarks, we will take the parameters of the planned new solenoid magnet at Oak Ridge National Laboratory. This 13~T, 50~cm bore magnet is expected to be received by the end of 2022, and would provide unparalleled sensitivity. 

The bore dimensions of the magnet provide one of the main volume constraints on the setup: as it is a warm bore magnet, we must leave some space for cryogenic infrastructure, which we will take as leaving a 45~cm useable bore at least 75~cm long. Through the tuning mechanism discussed in Section~\ref{sec:tuning} we expect 15-30\% tuning with minimal volume losses, meaning that for the full frequency range we would need to swap out inserts, with several inserts to cover a decade of parameter space. By swapping inserts we can also optimise the wire radius to keep an almost optimal quality factor for all frequencies.

A more stringent size limitation for higher frequencies would be the requirement for single-moded operation. While a multi-antenna setup could be employed, for a baseline design we will instead restrict the dimensions of the device to be less than $\sqrt{Q_{U}}\lambda_c$. One must also optimise the geometry of the system. For efficiently filling a solenoid, a cylindrical geometry is best, though this increases the probability of parasitic modes in the system if air gaps large enough to support modes are created. %A more conservative square design would largely avoid these design challenges, though reduces the used volume by 40\%.
This consideration would allow us to use a single port, coupled with for example a monopole antenna, to read out into a heterodyne receiver. For a baseline we will consider commercially available high electron mobility transistors (HEMTs) which typically get within a factor of a few of the standard quantum limit (SQL). As an upgrade path more advanced quantum readouts will be considered, which have been shown to reach or beat the SQL, as demonstrated in HAYSTAC~\cite{HAYSTAC:2020kwv}.

One potential issue is the presence of many TE modes near the target frequency if a simple metallic cavity is used. If mode mixing between TE and TM modes becomes an issue, there are a number of ways to ameliorate it. If the plasma haloscope is sufficiently large, radiative losses may be subdominant to resistive wire losses and no cavity is needed. However, even in the case that a cavity is needed, transverse wires could be implemented to raise the TE frequencies and reduce or eliminate the modes near the plasma frequency of interest. Alternatively, the metallic cavity walls could be replaced with a photonic band gap (PBG) cavity, which would not form standing TE modes. Depending on whether regular cavity walls, absorber or a PBG surrounds the metamaterial, the active volume of the system will change. The smallest volume would occur for a PBG, leaving an approximately 35~cm usable bore. 

As shown in Ref.~\cite{Balafendiev:2022wua} the primary source of losses comes from the resistivity of the wires. We will take the cryogenic quality factor predictions shown in Fig.~\ref{fig:radius} as a guide for the unloaded quality factor of the full experiment. However, superconducting wires would allow for significantly higher quality factors to be reached, which may be particularly beneficial in the high-mass parameter space. Recently, the idea of using dielectric rod inserts to form a kind of photonic crystal has been proposed, which may also provide an avenue for higher quality factors~\cite{Bae:2022ydq}.  

The design of ALPHA laid out in this section was chosen to minimise risks, taking the simplest read out method and most promising tuning methods. However, any new approach of course requires validation and contains some potential technical challenges or risks. The tuning method, while demonstrated numerically in Section~\ref{sec:numerical}, requires a practical demonstration in a prototype. Further, any tuning system comes with the risk of mode mixing or crossings when tuning, which may require mitigation or avoiding certain frequencies. Further, while the prototypes have shown good agreement with theory at different physical scales (for example, 10 and 40 layers, both inside and outside of a cavity) scaling up to a full experiment may lead to new challenges. This said, the current experimental and numerical results support the ALPHA design as one of the most promising avenues to pursue high mass axions.
\section{Discovery Potential \label{sec:projected}}
With these design considerations, we can estimate the potential sensitivity of the ALPHA experiment. We will consider several stages for the experiment, corresponding to upgrades in detector technology to allow for higher frequency parameter space to be explored to smaller couplings. We will consider a two stage design for ALPHA, the first operating with commercially available technology (ALPHA Stage I) and the second using an upgraded detector design reaching the SQL (ALPHA Stage II).

The projections are computed assuming a setup
as close as possible to the actual implementation
and similar to analogous experiments --- see e.g., Refs. 
\cite{Brubaker:2017rna,Brubaker:2017ohw}.
In general, a measurement campaign would consist
of measuring at a given frequency $\nu_r$ for a time
$\tau_r$, tuning to the next frequency and then
repeating, until the desired frequency range is
covered or the expected livetime $\mathrm{T} =
\sum_r\tau_r$ is reached.
If an excess is measured above a certain threshold then a rescan will be triggered to remeasure that frequency either
confirming or rejecting a signal candidate. In addition, case must be taken to ensure maximum sensitivity to HP dark matter~\cite{OHare:2021zrq}. As it is possible that HP dark matter may have a fixed polarisation ($E$-field direction) over astrophysically relevant scales, in order to ensure a potential signal is not missed each frequency should be measured at three different times (separated by several sidereal hours~\cite{OHare:2021zrq}).
For the purpose of of a simple estimate of the discovery potential we can just set the rescan trigger threshold to be sufficiently high that the time spent rescanning does not affect the overall scan time. Thus we will focus on the statistics without rescans; for a rigorous method for statistical inference with rescanning we refer the reader to Ref.~\cite{GRACStatPaper}.

In our analysis framework, we consider the range over which our
experiment is sensitive to be covered by a collection of
$N_\mathrm{sp}$ noise spectra, each different by a sequential
tuning step $\nu_{r}$ of the resonance ($r=1,\dots,\,N_\mathrm{sp}$).
For each measurement step $\nu_{r}$, we assume to measure for a time
$\tau_r$ and gather a spectrum of width $2\Delta\nu_r$.
Each spectrum is broken down into a number $N_{r}$ of bins
$\nu_{rb}$ ($b=1,\dots,\,N_{r}$) giving a high frequency resolution.
Each bin stores the noise fluctuations in the bin range
$\Delta\nu_{rb}$. The time resolution required is given by the inverse of the bin width, $\tau_{rb} = \Delta\nu_{rb}^{-1}$ and averaged over
the total acquisition time $\tau_r$.
%\Alex{What exactly does this mean? is this needed?}.

Under this assumptions, and thanks to the central limit theorem,
the noise fluctuations normalized with respect to the
spectral baseline, $w_{rb}$, are expected to be normally distributed.
Each one of the $w_{rb}$ can be regarded as a random variable
with a standard deviation $\sigma_{rb} =1/\sqrt{\tau_r\Delta\nu_{rb}}$
and zero mean. On the other hand, if an axion happens to deposit power
in a given bin, the mean $\mu_{rb}$ for a frequency $\nu_a$, bin $b$ and and spectrum $r$ will be shifted from zero by a factor:
%\Alex{Can we clarify the different "means" and their meaning? Also $P_s$ already has factors of $C_{a\gamma}$. }
% \begin{equation}
% \mu_{a,rb} = C_{a\gamma}^2 u_{a,rb}\,,
% \end{equation}
% where $u_{a,rb}$ embodies the dependency on the axion frequency
% $\nu_a$, the bin $b$ and the spectrum $r$:
\begin{align}
	%\mu_{a} = C_{a\gamma}^2\times
	\mu_{rb} &=
	\frac{P_s(\nu_a,\,\text{param.})}%
	{T_{\mathrm{sys}}(\nu_r,\,\nu_{rb})}
	\sqrt{\frac{\tau_{rb}}{\Delta\nu_{rb}}}\,
	D(\nu_a,\,\nu_r)\,L(\nu_a,\,\nu_{rb})\,,\nonumber \\
 &\equiv C_{a\gamma}^2 u_{rb}\,,
\end{align}
where $T_{\rm sys}$ is the system noise temperature and we have defined $u_{rb}$ to factor out the coupling constant on which limit will be placed.\footnote{Here for simplicity we focus on the case of an axion, however a HP signal has an analogous expression simply replacing $P_s$ with $P_s^{\rm HP}$ and $C_{a\gamma}$ with $\chi$ as defined in Eqn.~\eqref{eq:angle}.}
%as expressed in Eqn.\eqref{eq:power}.
As the measured frequency $\nu_r$ may not exactly match the axion frequency $\nu_a$ we include a factor $D(\nu_a,\,\nu_r)$ \cite{Brubaker:2017rna,Brubaker:2017ohw}
\begin{equation}
    D(\nu_a,\,\nu_r) \sim
    \left[1+\left(2\frac{\nu_{a}-\nu_{r}}{\Delta\nu_{r}}
	\right)^2\right]^{-1}.
\end{equation}
Similarly, $L(\nu_a,\,\nu_{rb})$ parameterizes the amount of signal
falling in the $b$-th bin of the $r$-th spectrum, when
$\Delta\nu_{rb}$ is smaller than the signal width 
$\Delta\nu_a\simeq 10^{-6}\nu_a$ \cite{Brubaker:2017rna,Brubaker:2017ohw}
\begin{equation}
    L(\nu_a,\,\nu_b) = \int_{\nu_{rb}}^{\nu_{rb}+\Delta\nu_{rb}}
    f(\nu|\nu_a)\,\text{d}\nu,
\end{equation}
where $f(\nu|\nu_a)$ is the axion kinetic energy distribution, which we take to be Maxwellian.

To evaluate whether a measured value gives evidence for the presence of
dark matter we must test the signal hypothesis at given frequency $\nu_a$.
To do so we use a signed-root likelihood ratio test~\cite{jensen}
$s=w/u$
%\begin{equation}
%    s = \frac{w}{u},
%\end{equation}
where $w$ and $u$ are quantities built from
the relevant bins to test the hypothesis.
That is,
\begin{equation}\label{eq:defUkXk}
    w = \sum_{r}\sum_{b}
    \frac{u_{rb}w_{rb}}{\sigma_{rb}^2}
    \quad\text{and}\quad
    u = \sum_{r}\sum_{b}
    \frac{u_{rb}^2}{\sigma_{rb}^2},
\end{equation}
with the sum extended over the bins $\nu_{rb}$ affected by a possible
signal, i.e., within the axion line width
$\nu_a\leq\nu_{rb}\leq\nu_a + \Delta\nu_a$.
%\begin{equation}
%    \nu_a\leq\nu_{rb}\leq\nu_a + \Delta\nu_a
%\end{equation}
%$\nu_a\leq\nu_{rb}\leq\nu_a + \Delta\nu_a$.  

The test statistics $s$
follows a standard normal distribution under
the null hypothesis; that is, in the absence of
any signal. If the
value of the test statistic inferred
from the data $\hat{s}$, lies in the tail of the distribution
beyond the $5\sigma$ quantile the
discrepancy with the background-only hypothesis
is deemed significant and a discovery can
be claimed. The coupling constant $\left|C_{a\gamma}\right|$
such that the distribution of $s$ is above $\hat{s}$ with
a significance $\alpha$ is given by \cite{GRACStatPaper}
\begin{equation}
    \left|C_{a\gamma}\right|^2 = \frac{\hat{s} -\Phi^{-1}(\alpha)}{\sqrt{u}},
\end{equation}
where $\Phi^{-1}$ is the cumulative distribution of a 
standard normal. In most high energy physics the convention
is to make discovery projections assuming that the assumed coupling
constant triggers a discovery (globally) 50\% of the time.
For our purposes in this paper, we quote the value
of $|C_{a\gamma}|$ having 50\% probability to
exceed a 5$\sigma$ local threshold. That is,
\begin{equation}
    \left|C_{a\gamma}\right| =
    \sqrt{\frac{5}{\sqrt{u}}}.
\end{equation}
\begin{figure*}[t]
\includegraphics[width=14cm]{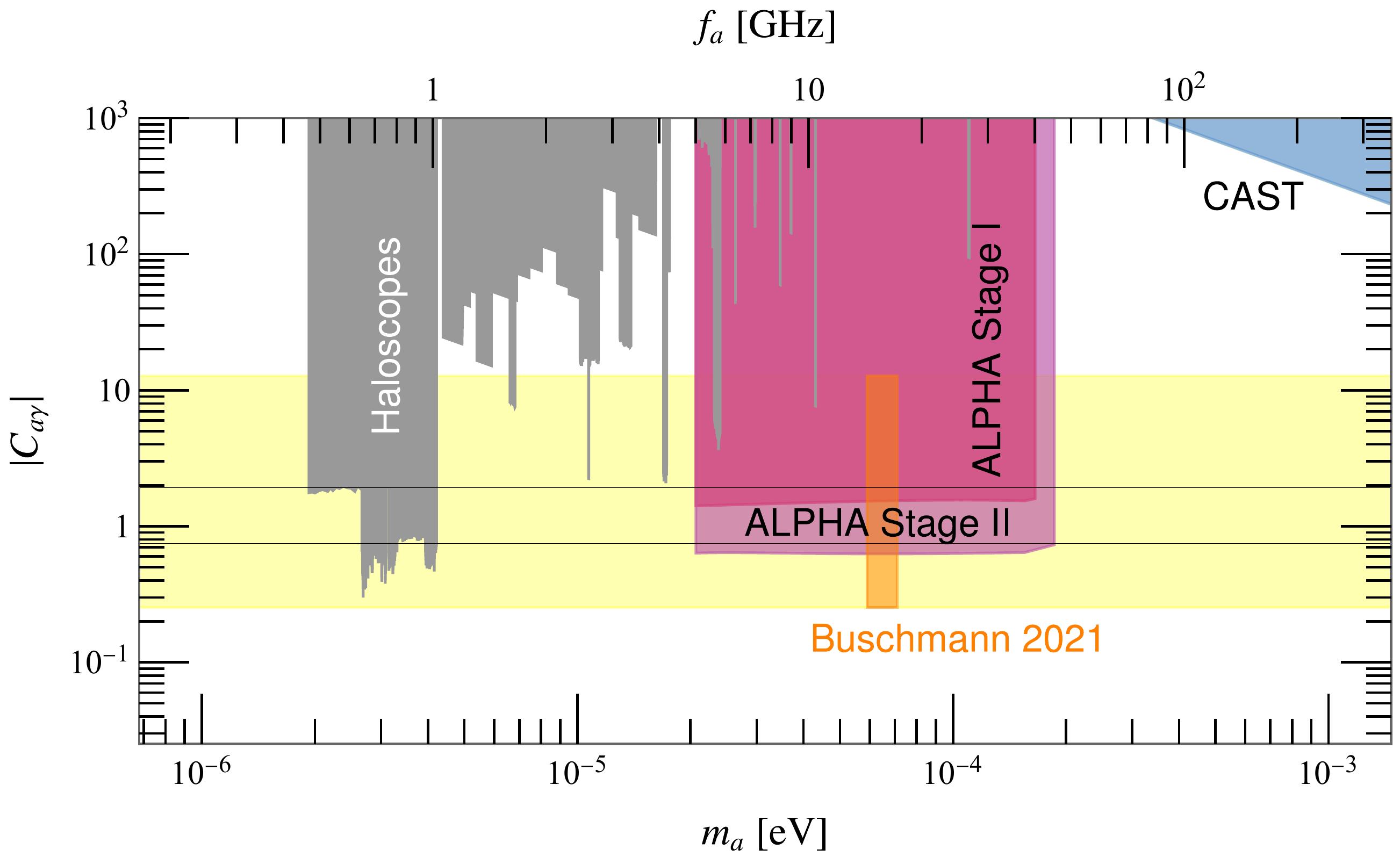} 
  \caption{In red we show the discovery potential for ALPHA for axions in the $\{m_a,|C_{a\gamma}|\}$ plane of a plasma with an unloaded quality factor given by Fig.~\ref{fig:radius} inside a $13\,$T magnetic field with effective radius 35~cm and bore length 75\,cm. We assume a two stage experiment, with both systems critically coupled. In ALPHA Stage I we assume a 2-year data taking time with commercially-available HEMT amplifiers and will cover to the KSVZ axion for $5-40$\,GHz. ALPHA Stage II will use a two year quantum-limited detection campaign to cover $5-45\,$GHz down to DFSZ. Note that the projected exclusions are better by a factor of 0.64, with even Stage I being able to exclude axions at 95\% confidence level down to $|C_{a\gamma}|\simeq 0.9$ The black lines bracket the traditional axion model band, $0.746<|C_{a\gamma}|<1.92$, with the yellow band showing the extended QCD axion model band \cite{DiLuzio:2020wdo}. Existing haloscope exclusion limits are displayed in gray~\cite{DePanfilis:1987dk,Hagmann:1990tj,Asztalos:2001jk,Asztalos:2009yp,CAPP:2020utb,McAllister:2017lkb,Du:2018uak,Zhong:2018rsr,Lee:2020cfj,Alesini:2020vny,Melcon:2021dyi,Backes:2020ajv}, with the CAST limit shown in blue~\cite{Anastassopoulos:2017ftl}. The most recent prediction for the axion mass in a post inflationary scenario assuming a scale invariant spectrum is shown in orange~\cite{Buschmann:2021sdq}. }
     \label{fig:reach} 
 \end{figure*}

\begin{figure*}[t]
\includegraphics[width=14cm]{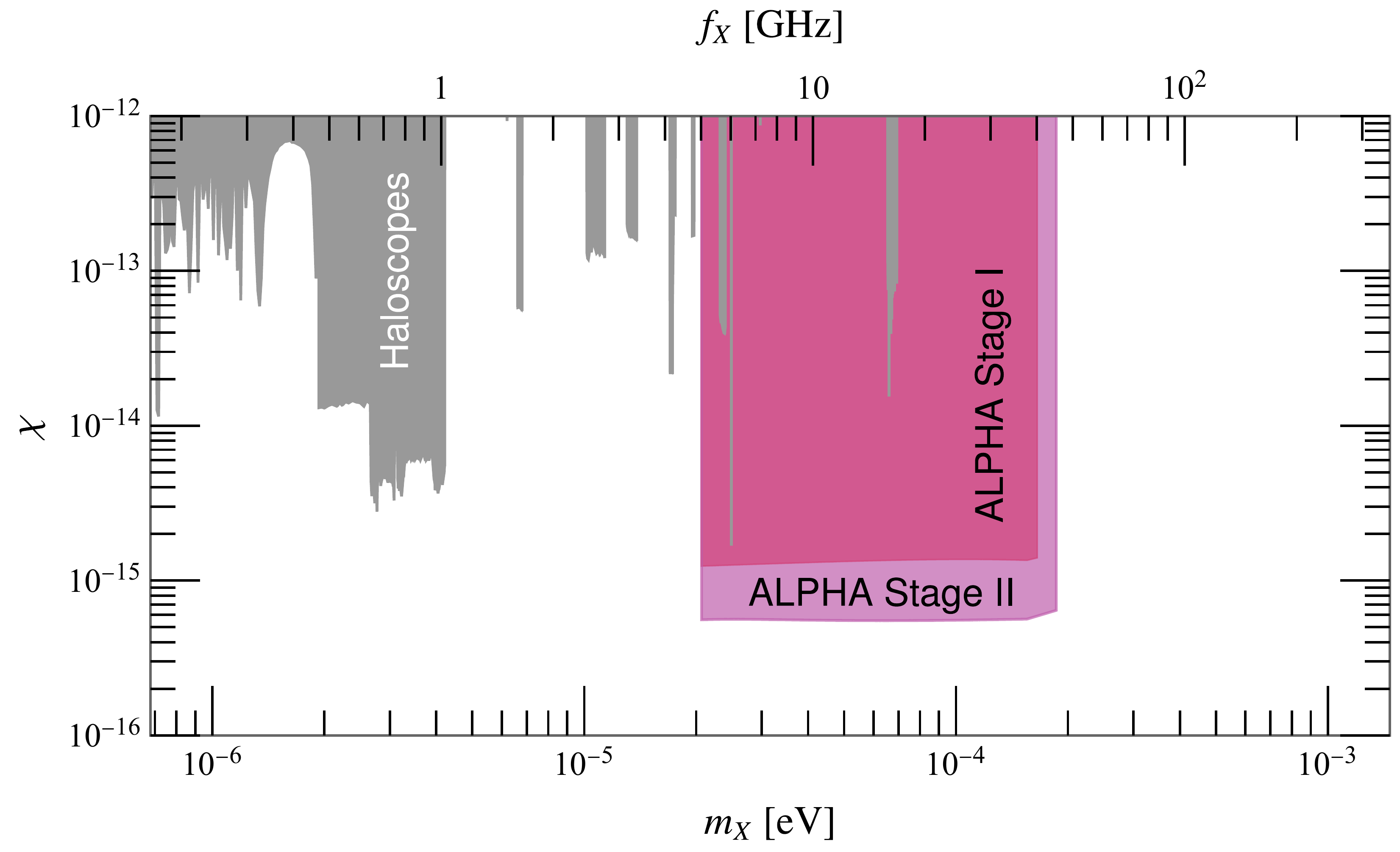} 
  \caption{In red we show the discovery potential for ALPHA for hidden photons in the $\{m_X,\chi\}$ plane of a plasma with an unloaded quality factor given by Fig.~\ref{fig:radius} with a radius of 35~cm and length 75\,cm. We assume a two stage experiment, with both systems critically coupled. In ALPHA Stage I we assume a 2-year data taking time with commercially available HEMT amplifiers operating between $5-40$\,GHz. ALPHA Stage II will use a two year quantum-limited detection campaign to cover $5-45\,$GHz. We assume the conservative fixed polarisation scenario, as outlined in Ref.~\cite{Caputo:2021eaa}. To avoid the possibility of missing a discovery of dark matter ALPHA will combine multiple shorter measurements throughout the sidereal day. Existing limits are displayed in gray~\cite{Asztalos:2009yp,Du:2018uak,Boutan:2018uoc,Zhong:2018rsr,Braine:2019fqb,Nguyen:2019xuh,Backes:2020ajv,Lee:2020cfj,Jeong:2020cwz,Kwon:2020sav,Alesini:2020vny,Dixit:2020ymh,Godfrey:2021tvs,Cervantes:2022yzp}, using the rescaling suggested in Ref.~\cite{Caputo:2021eaa} and expanded with up to date experiments in Ref.~\cite{ciaran} .  }
     \label{fig:reachHP} 
 \end{figure*}

That would label the given fluctuation as a promising
candidate and trigger deeper
investigations, aiming to reach the global
significance needed for claiming a discovery.
Assuming that the step size for testing for a potential signal is a
small fraction of the axion linewidth (in our case explicitly
$\Delta\nu_a/5$ to avoid missing a potential signal), such a
method would lead to $\sim 3$ rescans over a decade of
parameter space, which would not
affect our time estimates. The search can be further enhanced by optimising the time spent in the initial scan period vs rescanning for possible signals by adjusting the rescan threshold.

While our procedure gives numerically similar results to that described in Refs.~\cite{Lawson2019,Gelmini:2020kcu}, it has both a greater formal meaning and is much more easily extendable to realistic data analysis. In Refs.~\cite{Lawson2019,Gelmini:2020kcu} they simply integrated the Dicke radiometer equation over the frequency range an required a signal-to-noise ratio of four (approximating the frequency response as rectangular rather than Lorenzian). However, this simpler approach lacks a rigorous statistical interpretation in terms of the probability of correctly making a discovery. 

To make some specific predictions, we will conservatively take the volume of a PGB bounded metamaterial, and use the quality factors calculated in Fig.~\ref{fig:radius}.  As discussed in Section.~\ref{sec:design}, we take a 13\,T magnet with 35\,cm useable bore diameter and 75\,cm long. We further restrict the system at high frequencies to be less than $\sqrt{Q_U}\lambda_c$ in any dimension to maintain a simple single antenna readout. Commercially available HEMTs currently can provide a noise temperature of $\sim 4\,$K below 20\,GHz and $\sim 8\,$K below 40 GHz.\footnote{For an example, see \href{www.lownoisefactory.com}{www.lownoisefactory.com}.} We assume that ALPHA Stage I has two years of effective data taking time, followed by two years in Stage II at the SQL $T=\omega_a$.

We show the discovery potential of ALPHA in Fig.~\ref{fig:reach}. 
Note that, in case no signal is detected, the projection for the 95\%
confidence-level upper limit\footnote{That is, the coupling constant that
would give a result equally unlikely 5\% of the time.} would be
scaled of a factor $0.64$.
In Stage I, ALPHA would be capable of discovering KSZV axions from
$5-40$ GHz, with projected exclusions at 95\% CL extending down to $|C_{a\gamma}|\simeq 0.9$. Under the assumption of an upgraded amplifier, Stage II reaches the DFSZ axion  over $5-45$\,GHz, with an expected 95\% CL exclusion limit of $|C_{a\gamma}|\simeq0.35$. Above 40\,GHz the sensitivity decreases due to our assumption of a single-moded readout (in other words, the volume of the haloscope is reduced to avoid having to use multiple antennas). However, using multiple antennas would extend the DFSZ reach up to 50\,GHz.
 
 These projections demonstrate similar detection potential to the proposed MADMAX experiment~\cite{MADMAX:2019pub}, a dielectric haloscope planned to operate in a similar frequency range. We caution that projections should be used to give an idea of the frequency range and requirements of a full scale experiment, rather than as a comparative tool. As the methodologies and assumptions of various experiments' predictions are different, as well as the experiments having different engineering challenges and infrastructure considerations, it is difficult to make a fair comparison. However, as the high frequency range has been historically very experimentally challenging, it is important to have multiple complementary approaches in order to ensure that the full range of axion masses is searched.

In addition to performing highly sensitive axion searches, ALPHA will also look for HP DM. While the electromangetic response is similar for both cases (i.e., the same mode will be excited in the haloscope, and be read out in the same way), HP DM may have a non-trivial polarisation. As shown in Eq.~\eqref{eq:angle}, the signal power depends on the angle between the wires and HP polarisation $\theta$, integrated over the course of the measurement (or measurements, if multiple are done on the same frequency). 

If the polarisation of the HP is random, so that at every coherence time a different polarisation is sampled, we are left with a simple average $\langle \cos^2\theta\rangle_T=1/3$. However, depending on the state of the HP polarisation, $\langle \cos^2\theta\rangle_T$ can have a highly non-trivial dependence on the measurement time and method. In particular, the most conservative assumption is that the polarisation is fixed over scales much larger than the laboratory, leading to the possibility that the signal can be missed in short measurements due to misalignment with respect to the sensitive axis of the experiment~\cite{Arias:2012az}. To overcome this, it was proposed in Ref.~\cite{Caputo:2021eaa} that combining multiple shorter measurements taken at different sidereal times can eliminate this possibility without affecting an axion search, giving $\langle \cos^2\theta\rangle_T\simeq 1/3$. ALPHA plans to employ such a strategy, as well as performing dedicated HP analysis in order to maintain maximum HP sensitivity. As such a measurement can be with the same data as with an axion search, we show the expected discovery potential for HP in Fig.~\ref{fig:reachHP} in the conservative fixed polarization scenario.

 As we can see from this analysis, with conservative assumptions regarding volume and quality factors ALPHA has a high discovery potential for axions and HP DM over a wide range of frequencies ($5-45$\,GHz). In particular, Stage I can discover KSVZ axions with two years of livetime, during which quantum-limited detectors will be explored. This will lead to Stage II, pushing down to DFSZ over a two year campaign. 
 
 \ \ \ \ \ \ \ \ \ \ \ \

\section{Summary and Conclusion\label{sec:conclusion}}
The work reported here extends, refines, tests, and validates the broad conceptual design for an axion haloscope exploiting plasma resonance in a wire metamaterial proposed in Ref.~\cite{Lawson2019}. Numerical simulations (Section \ref{sec:numerical}) and experimental measurements using prototypes (Section \ref{sec:experiment}) support the conclusion that large resonators based on wire metamaterials can achieve quality factors (Q values) that are significantly larger than were assumed in Ref.~\cite{Lawson2019}; that appropriate levels of tunability can be achieved by mechanical adjustment of the wires (Section IIIC, Section IVB); and that a signal of the kind and magnitude predicted to be generated by a cosmological axion background can be detected in a low-noise environment (Section VI, Section VII).   Estimates of the potential sensitivity of devices based on these principles, using conservative extrapolations of this work and assuming practically achievable settings for volume and magnetic field strength, appear in Figure~\ref{fig:reach} and its caption.  Alternative read-out schemes and possible use of superconducting wires (Section V) are under active consideration.  Either of those developments might yield further improvements in sensitivity, as also would the use of larger volumes or higher field strengths.  

\section*{Acknowledgments}
 This project has received funding from the European Research Council (ERC) under the European Union’s Horizon 2020 research and innovation programme (grant agreement no. 101018897 CosmicExplorer) as well as funding for A.J.M and F.W under grant No. 742104. The work was also supported by the Swedish Research Council (VR) under Dnr 2019-02337 ``Detecting Axion Dark Matter In The Sky And In The Lab'' (AxionDM), with additional support for F.W under Contract No. 335-2014-7424. FW is also supported by the U.S. Department of Energy under grant Contract Number DE-SC0012567. The work of H.V.P was additionally supported by the G\"{o}ran Gustafsson Foundation for Research in Natural Sciences and Medicine. A.G.R. and J.C thank support Knut and Alice Wallenberg Foundation: ``Discovering Dark Matter Particles in the Laboratory". Fermilab is operated by Fermi Research Alliance, LLC under Contract No. DE-AC02-07CH11359 with the United States Department of Energy. The University of California Berkeley acknowledges support from NSF under grant number PHY-1914199. R.B. and P.B. were supported by Priority 2030 Federal Academic Leadership Program. S.M.A. acknowledges support from NSF under grant DMR-2004386, and the US Department of Energy under grant DESC0017931.
This manuscript has been supported by UT-Battelle, LLC under Contract No.~DE-AC05-00OR22725 with the U.S. Department of Energy as well as by the U.S. Department of Energy through the Oak Ridge National Laboratory LDRD Program.

\providecommand{\href}[2]{#2}\begingroup\raggedright\endgroup

\end{document}